\documentclass[12pt,amssymb]{article} 

\usepackage{mathrsfs}
\usepackage{amssymb}
\usepackage[dvips]{graphicx}

\newcommand{\newsection}[1]{
\addtocounter{section}{1} \setcounter{equation}{0}
\setcounter{subsection}{0} \addcontentsline{toc}{section}{\protect
\numberline{\arabic{section}}{{\rm #1}}} \vglue .6cm \pagebreak[3]
\noindent{ \bf  \thesection. #1}\nopagebreak[4]\par\vskip .3cm}
\newcommand{\newsubsection}[1]{
\addtocounter{subsection}{1}\setcounter{subsubsection}{0}
\addcontentsline{toc}{subsection}{\protect
\numberline{\arabic{section}.\arabic{subsection}}{#1}} \vglue .4cm
\pagebreak[3] \noindent{\it \thesubsection.
#1}\nopagebreak[4]\par\vskip .3cm}

\makeatletter
\newcommand{\seclabel}[1]{%
  \@bsphack
  \protected@write\@auxout{}%
     {\string\newlabel{#1}{{\thesection}{\thepage}}}
  \@esphack
  }
\newcommand{\subseclabel}[1]{%
  \@bsphack
  \protected@write\@auxout{}%
     {\string\newlabel{#1}{{\thesubsection}{\thepage}}}
  \@esphack
  }
\newcommand{\tablabel}[1]{%
  \@bsphack
  \protected@write\@auxout{}%
     {\string\newlabel{#1}{{\arabic{tabnum}}{\thepage}}}
  \@esphack
  }
\makeatother

\renewcommand{\theequation}{\thesection.\arabic{equation}}

\newlength{\extraspace}
\newlength{\extraspaces}
\newcounter{dummy}
\newcommand{\bc}{\begin{center}}
\newcommand{\ec}{\end{center}}
\newcommand{\be}{\begin{equation}
\addtolength{\abovedisplayskip}{\extraspaces}
\addtolength{\belowdisplayskip}{\extraspaces}
\addtolength{\abovedisplayshortskip}{\extraspace}
\addtolength{\belowdisplayshortskip}{\extraspace}}
\newcommand{\ee}{\end{equation}}

\newcommand{\ba}{\begin{eqnarray}
\addtolength{\abovedisplayskip}{\extraspaces}
\addtolength{\belowdisplayskip}{\extraspaces}
\addtolength{\abovedisplayshortskip}{\extraspace}
\addtolength{\belowdisplayshortskip}{\extraspace}}
\newcommand{\ea}{\end{eqnarray}}

\newcommand{\ban}{\begin{eqnarray*}
\addtolength{\abovedisplayskip}{\extraspaces}
\addtolength{\belowdisplayskip}{\extraspaces}
\addtolength{\abovedisplayshortskip}{\extraspace}
\addtolength{\belowdisplayshortskip}{\extraspace}}
\newcommand{\ean}{\end{eqnarray*}}
\newcommand{\baa}{
\addtocounter{equation}{1} \setcounter{dummy}{\value{equation}}
\setcounter{equation}{0}
\renewcommand{\theequation}{\thesection.\arabic{dummy}\alph{equation}}
\begin{eqnarray}
\addtolength{\abovedisplayskip}{\extraspaces}
\addtolength{\belowdisplayskip}{\extraspaces}
\addtolength{\abovedisplayshortskip}{\extraspace}
\addtolength{\belowdisplayshortskip}{\extraspace}}
\newcommand{\eaa}{
\end{eqnarray}
\setcounter{equation}{\value{dummy}}
\renewcommand{\theequation}{\thesection.\arabic{equation}}}

\input epsf

\newcounter{fignum}
\newcounter{tabel}

\newcounter{tabnum}
\newcommand{\vev}[1]{\left\langle #1\right\rangle}

\newcommand{\half}{\frac{1}{2}}
\newcommand{\del}{\partial}
\newcommand{\delb}{\bar{\del}}
\newcommand{\eol}{\nonumber \\}

\newcommand{\cO}{{\cal O}}
\newcommand{\Ext}{{\rm Ext}}
\newcommand{\Hom}{{\rm Hom}}

\newcommand{\Gr}{{\rm Gr}}
\newcommand{\W}{{\rm W}}
\newcommand{\MHS}{{\rm MHS}}

\begin{document}

%
%

\begin{flushright}
November 2012\\
\end{flushright}
\vspace{2cm}

\thispagestyle{empty}

\begin{center}
{\Large\bf Weak Coupling, Degeneration and \\
Log Calabi-Yau Spaces
 \\[13mm] }

{\sc Ron Donagi}\\[2.5mm]
{\it Department of Mathematics, University of Pennsylvania \\
Philadelphia, PA 19104-6395, USA}\\[5mm]

{\sc Sheldon Katz}\\[2.5mm]
{\it Department of Mathematics, University of Illinois\\
1409 W.~Green St., Urbana, IL 61801, USA}
\\[5mm]

{\sc Martijn Wijnholt}\\[2.5mm]
{\it Arnold Sommerfeld Center, Ludwig-Maximilians Universit\"at\\
Theresienstrasse 37 \\
D-80333 M\"unchen, Germany }\\
[15mm]

Abstract:
\end{center}

We establish a new weak coupling limit in $F$-theory. The new
limit may be thought of as the process in which a local model
bubbles off from the rest of the Calabi-Yau. The construction
comes with a small deformation parameter $t$ such that
computations in the local model become exact as $t \to 0$. More
generally, we advocate a modular approach where compact Calabi-Yau
geometries are obtained by gluing together local pieces (log
Calabi-Yau spaces) into a normal crossing variety and smoothing,
in analogy with a similar cutting and gluing approach to
topological field theories. We further argue for a holographic
relation between $F$-theory on a degenerate Calabi-Yau and a dual
theory on its boundary, which fits nicely with the gluing
construction.

\newpage

\renewcommand{\Large}{\normalsize}

\tableofcontents

\newpage

\newsection{Introduction}

In type II string theories, gauge and matter fields are localized on
submanifolds of the ten-dimensional space-time. Therefore the study
of type II compactifications (or their $M$/$F$-theory versions)
usually proceeds by splitting the analysis in two steps: first one
studies a local model in detail, and then one tries to embed it in a
global UV completion. Furthermore, it has been suggested that the
hierarchy between the weak scale and the Planck scale, or between
the GUT scale and the Planck scale, make such a two-step procedure
natural. This raises the important question: what is the precise
relationship between the local model and the global model?

One expects that a local model will arise as a limit of a global
model. In \cite{Verlinde:2005jr} a decoupling limit was defined by
taking the ratio of the volume of the divisor supporting the
Standard Model and the volume of the total space to zero. In
$F$-theory, this was studied in detail in
\cite{Donagi:2009ra,Cordova:2009fg}. However, this is a limit in
the K\"ahler moduli space, which has no effect on the complex
structure of a global model. In the holomorphic world, functions
are determined by their behaviour on a small neighbourhood, so
`infinity does not decouple.'

We can also see this at the level of the effective action. The
$4d$ $N=1$ supergravity is characterized by a K\"ahler potential
${\cal K}$, a superpotential $W$ and a gauge kinetic function $f$.
Non-renormalization theorems restrict the dependence of $W$ and
$f$ on the K\"ahler moduli. As a result, they are largely
unchanged in the limit above, and remember the structure of the
global model.

In the present paper, we will address this question more
systematically for $F$-theory models, although our ideas should
also apply to other settings, such as the situation considered in
\cite{Verlinde:2005jr}. From the above discussion it is clear that
we need to take a limit to a boundary of the complex structure
moduli space. But this must be done carefully, for otherwise the
relation between the local and the global model will be very hard
to understand. We will see that one can bubble off local models
from global models by semi-stable degeneration. In particular, we
will define a new type of semi-stable degeneration in $F$-theory
which recovers the local models studied in \cite{Donagi:2009ra}
from a global model. (For earlier work, see
\cite{Donagi:2008ca,Beasley:2008dc,Hayashi:2008ba}.) If the global
model admits a K3 fibration, then we recover a known semi-stable
degeneration of the K3 fibers. However, our method also applies if
there is no K3 fibration.

The main purpose of this paper however is to place the results of
section \ref{localglobal} and reference \cite{CDW} in a broader
perspective. Namely we would like to discuss the roles played by
degeneration and its converse, which is a nice way to construct
global models from local pieces, as a cutting-and-gluing procedure
in holomorphic field theories.

To explain this, let us first define the notion of a log
Calabi-Yau space, since this has not yet percolated to the physics
literature and it is one of the fundamental concepts studied in
this paper. We define a log Calabi-Yau to be a pair $(X,D)$, where
$X$ is a variety and $D$ is an effective divisor in $X$ such that
the log canonical class $K_{(X,D)} \equiv K_X + D$ vanishes. It
admits a unique $(n,0)$-form which is is holomorphic on
$X\backslash D$ and has a logarithmic pole along $D$, whose
residue is the holomorphic $(n-1)$-form of $D$. A log Calabi-Yau
should be thought of as the holomorphic analogue of a real and
oriented manifold with boundary.

Now we start with a disconnected sum of Calabi-Yau manifolds with
boundary, say $(X_1,D)$ and $(X_2,D)$ in the simplest case.  Then we
glue these pieces together along the boundaries into a normal
crossing variety
\be Y_0 \ = \ X_1 \cup_D X_2. \ee
Finally when $Y_0$ has a log structure and satisfies other mild
hypotheses, we can deform $Y_0$ to a smooth global model $Y_t$,
where $t \in {\bf C}$ is a parameter, by the fundamental smoothing
theorem of Kawamata and Namikawa \cite{KawNam}. Obviously this
construction comes with a small parameter and a built-in
degeneration limit.

In order to make use of this technique in $F$-theory, it is
crucial to know that when $Y_0$ has an elliptic fibration, this
continues to hold for the smoothings $Y_t$.  We posed this as an
open problem at a recent meeting on string theory for
mathematicians at SCGP. To our delight, J\'anos Koll\'ar has now
settled this in the affirmative and proved a number of additional
results in \cite{KollarEllipticReview}. We include a discussion of
some of his results in section \ref{EllipticCriteria}.

Now for physics purposes we are not just interested in complex
manifolds per se, but in holomorphic field theory on such space,
that is a theory which depends on the complex structure but not on
the K\"ahler moduli. So in order to use surgery ideas, we want to
know how a holomorphic field theory behaves on a family $Y_t$. We
expect that its partition function and correlators behave nicely
in a semi-stable degeneration limit, and that these quantities can
in fact be constructed explicitly from the limit as an expansion
in $t$.

Furthermore, we expect an interesting interplay between the theory
in the bulk and the theory on the boundary.  In particular for the
case of degeneration in $F$-theory, we will argue in sections
\ref{CYGluing} and \ref{VHS} that computations can also be
performed `holographically' in the boundary theory. We discuss
some expected features of the boundary theory and how various
known examples fit the pattern. We believe these points give a new
perspective on compactifications in which log varieties play a
central role, for $F$-theory and more generally for string or
field theories on complex spaces.

\begin{figure}[t]
\begin{center}
            \scalebox{.5}{
  \includegraphics[width=\textwidth]{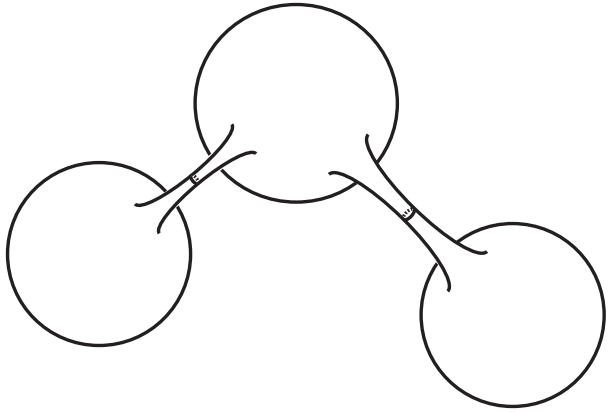}
}
\end{center}
 \caption{Gluing a Calabi-Yau manifold from local pieces.} \label{GluedCY}
\end{figure}

Coming back to $F$-theory, recall that for $F$-theory
compactifications to eight dimensions there are essentially only
two weak coupling limits (more precisely, type II degenerations
\cite{Clingher:2003ui}), which yield the $E_8\times E_8$ and the
$SO(32)$ heterotic string. As will hopefully be clear, below eight
dimensions there seems to be a veritable zoo of new degeneration
limits, and the methods discussed in this paper allow one to
construct them. The global-to-local degeneration which splits off
a local $dP_9$-fibration will be our main example in this paper,
but much more exotic examples could be considered. A closely
related paper \cite{CDW} examines another new type of
degeneration, namely a stable version of the Sen limit, from the
general point of view advocated here.

In section \ref{VHS} we discuss the $t$-dependence of the
Lagrangian in a degeneration limit by applying the theory of mixed
Hodge structures. Although somewhat abstract, this approach is
extremely powerful and allows us to fix much of the physics from
very little input. An interesting aspect of this analysis is the
interpretation of brane superpotentials as describing an
obstruction to splitting the mixed Hodge structure into pure
pieces.

Our approach answers several questions that were not clear
previously. Regarding the construction of phenomenological GUT
models along the lines of \cite{Donagi:2009ra}, we believe that
one of the most important issues that are clarified is that there
is a new small expansion parameter $t$ in the game, which we
identify explicitly, such that holomorphic calculations in the
local model become exact in the limit $t\to 0$. This explains to
what extent we can trust the $E_8$ picture of local models, and
makes the notion of `dropping the subleading terms' in
\cite{Donagi:2009ra} more precise.

We believe that having such a small modulus is an important step
forward in understanding global models, raising various
interesting phenomenological questions. Although the net amount of
chiral matter cannot change if we continuously vary $t$, more
detailed properties of the effective Lagrangian definitely depend
on $t$, as seen in section \ref{VHS}. The behaviour of the
Calabi-Yau metric as $t\to 0$ has not been much investigated, but
one expects an anisotropic picture as shown in figure
\ref{GluedCY}, where points on different local pieces get
infinitely separated in the limit. The modulus $t$ may also play a
role in the issue of sequestering.

\newpage

\newsection{Local and global models}

\seclabel{localglobal}

\newsubsection{Semi-stable degeneration}

\subseclabel{SemiStable}

The data of an $F$-theory compactification consists of an
elliptically fibered Calabi-Yau manifold with section, and a
configuration for a three-form field ${\sf C}_3$ with flux ${\sf
G}_4 = d{\sf C}_3$. In the present section, the focus will be on
aspects of the Calabi-Yau geometry. We will typically have in mind
elliptic fourfolds $\pi:Y \to B_3$, although the considerations
will be more general.

We would like to define global-to-local degenerations for
$F$-theory models. By this, we mean we would like to bubble off a
four-fold consisting of a $dP_9$-fibration over a divisor $S$ in
$B_3$, which should capture the behaviour of the elliptic
fibration $\pi:Y \to B_3$ along $S$. As in \cite{Donagi:2009ra},
we could do a further degeneration along the lines of \cite{DEL},
in order to relate the $dP_9$-fibration to a Hitchin system.

Let us consider a flat family $\pi_{\cal Y}: {\cal Y} \to \Delta$,
where
$\Delta$ is a disk parametrized by a variable $t$, such that
the generic fibers are smooth but the central fiber $\pi_{\cal
Y}^{-1}(0)$ is not. We will actually be interested in fibers with
certain types of singularities, which can however be resolved and
followed through the degeneration. This caveat will not affect
the validity of our techniques however.

In order to get a reasonable
limiting behaviour, the degeneration needs to be fairly mild.  Specifically,
we
will be able to achieve our objectives by a {\em semi-stable degeneration\/}.
This means that
the total space ${\cal Y}$ of the family is smooth and the central fiber is
reduced and has at most normal crossing singularities. In local
coordinates $(z_1, \ldots, , z_n)$, a normal crossing singularity
is a singularity given by an equation of the form $z_1 \cdots z_k
= 0$.

According to Mumford's semi-stable reduction theorem, any
degeneration may be put in the form
\be z_1 \cdots z_k \ = \ t \ee
for small $t$, possibly after a base change (i.e.\ pulling back
the base of the family by an analytic map $\Delta\to\Delta$)
and then blowing up and down. A
semi-stable degeneration need not be unique, as there may
be many different birationally equivalent models. A degeneration
is said to be {\em stable\/} if in addition, the central fiber has no
infinitesimal automorphisms.

Semi-stable degenerations have a number of nice properties. One of
these properties is that they essentially become `smooth' if we
include logarithmic differential forms. Let us denote the fibers
$\pi_{\cal Y}^{-1}(t)$ by $Y_t$, and the normal crossing divisor
of $Y_0$ by $Z$. Suppose that in local coordinates, the equation
of $Z$ is given by $z_1 \cdots z_k = 0$. Let us consider the
spaces
\be \Omega^q_{Y_0}(\log Z) \ee
of logarithmic differential forms on $Y_0$. For $q=1$, this is defined to
be the subsheaf of meromorphic differential forms on $Y_0$
generated by $dz_i/z_i$ for $1\leq i\leq k$ and $dz_{i}$ for
$k+1\leq i\leq n$, together with the relation
\be
{dz_1\over z_1} + \ldots + {dz_k\over z_k} \ =\ 0
\ee
For larger $q$ we just take exterior powers. More generally one
can define logarithmic forms with respect to a log structure,
but we will not do so here. The exterior derivative takes
logarithmic forms to logarithmic forms, so we get a complex and
we can compute cohomology. It will be convenient to use the
short-hand notation
\be H^k_{\rm log}(Y_0) \ \equiv \ {\mathbb H}^k(Y_0,
\Omega_{Y_0}^\bullet(\log Z)) \ee
for the logarithmic de Rham cohomology of $Y_0$. The spectral
sequence for ${\mathbb H}^k$ degenerates at $E_1$, so we get the
decomposition $H^k_{\rm log}(Y_0) =
\sum_{k=p+q}H^p(\Omega^q_{Y_0}(\log Z))$ on $Y_0$.  As motivation for
section~\ref{CYGluing}, suppose that $k=2$ and $W_1,\ W_2$ are the
components of $Y_0$.  Then the restriction to $W_i$
of $\Omega^n_{Y_0}(\log Z)$ is just the log canonical bundle $\Omega_{W_i}(Z)$
of the pair $(W_i,Z)$.  In particular if $Y_0$ is Calabi-Yau in the sense that
$\Omega^n_{Y_0}(\log Z)$ is trivial, then each $(W_i,Z)$ is log Calabi-Yau.

In a degeneration limit, the ordinary cohomology typically jumps, as some
classes on $Y_t$ will disappear from the cohomology, and some
others may appear. According to the work of Steenbrink \cite{Steenbrink76}, the
cohomology of logarithmic forms gives a geometric way to realize
the `nearby cocycles' on $Y_0$. Concretely, for a normal
crossing degeneration the sheaves of differential forms
$\Omega^q_{Y_t}$ naturally fit together into a locally free sheaf
$\Omega^q_{{\cal Y}/\Delta}(\log Y_0)$ of relative forms, whose
restriction to the central fiber yields $\Omega^q_{Y_0}(\log Z)$.
Similarly, the sheaves $R^p\pi_{{\cal Y}*}\Omega^q_{{\cal
Y}/\Delta}(\log Y_0)$ are locally free over the base, i.e. they
correspond to holomorphic vector bundles over the base. The fiber
over $t\not = 0$ is given by $H^p(\Omega^q_{Y_t})$ and the fiber
over $t=0$ is given by $H^p(\Omega^q_{Y_0}( \log Z))$. In
particular, the dimensions of the cohomology groups
$h^p(\Omega^q_{Y_0}(\log Z))$ coincide with the Hodge numbers
$h^{p,q}(Y_t)$ of the smooth fibers. As we will
see later, these results (and some additional properties discussed in
appendix \ref{MHS}) are very helpful for elucidating the
physics of a semi-stable degeneration.

We will also need to introduce the logarithmic tangent bundle,
$T(-\log Z)$, as the dual of the logarithmic cotangent
bundle $\Omega^1(\log Z)$. In the local coordinates above,
$T(-\log Z)$ is generated by $z_i \del/\del z_i$ for $i=1,\ldots
,k$ and $\del/\del z_i$ for $i=k+1, \ldots, n$, together with the
relation $\sum_{i=1}^k z_i \del/\del z_i = 0$.
The tangent space to the smoothing component of the deformation
space is given by $H^1(T(-\log Z))$ \cite{KatoLogSmooth}. By
contracting with the logarithmic $(n,0)$ form, we can relate this
to $H^1(\Omega^{n-1}(\log Z))$.

Another nice aspect of semi-stable degeneration is that the
asymptotic behaviour of the Hodge structure is rather well
understood. In particular, we can use the orbit theorems
\cite{SchmidOrbits} to qualitatively understand the asymptotic
behaviour of the Lagrangian in the degeneration limit, as we will
discuss later.

\newsubsection{Normal cone degeneration}

Let us first focus on the base $B_3$ without worrying
about the elliptic fibration. Suppose that the discriminant locus
contains a a smooth divisor $S \subset B_3$ of ADE singularities. We
would like to be able to `zoom in' on $S$, i.e. we would like to
be able to consider the first order neighbourhood of $S$ in $B_3$.
There is a construction in algebraic geometry which allows us to
do this, namely the degeneration to the normal cone. We will first
quickly review this construction and refer the interested reader to
\cite{FultonInt} for more details.
Then in the next subsection, we will show how
the elliptic fibration can be extended over the normal cone
degeneration of $B_3$, yielding a semi-stable degeneration of the
elliptic fourfold.

The deformation to the normal cone is
constructed as follows. We consider the constant family
\be {\cal B}\ =\ B_3 \times {\bf C} \ee
parametrized by $t\in {\bf C}$.
We have the projection $\pi_{\cal B}: {\cal
B} \to {\bf C}$ and all fibers are canonically isomorphic to $B_3$.
Then we blow up ${\cal B}$ along the subvariety
$S \times \{0\} \subset {\cal B}$ to obtain a variety $\tilde{\cal
B}$ with exceptional divisor $E\subset\tilde{\cal B}$, with $E$
isomorphic to the
projectivization ${\bf P}({\cal O}\oplus N_S)$ of the normal bundle $N_S$
of $S$ in $B_3$.
Via the projection $\pi_{\tilde{\cal B}}:\tilde{{\cal B}}\to
{\bf C}$, we get another family of varieties parametrized by
$t$.  Let $\tilde{B}\subset\tilde{{\cal B}}$ be the proper transform of $B_3
\times\{0\}$, which projects isomorphically to $B_3$ via the composite map
$\tilde{\cal B}\to {\cal B}\to B_3$, where the first map is the map
$\rho:\tilde{\cal B}\to {\cal B}$ which
blows down $E\subset\tilde{\cal B}$ to $S\times\{0\}\subset {\cal B}$
and the second map is the projection $\pi_1$
of ${\cal B}=B_3\times{\bf C}$ to its first factor $S$.

Let us consider the structure of the family $\tilde {\cal B}$ in more
detail. Obviously for
$t\not = 0$, the fiber $\pi_{\tilde{\cal B}}^{-1}(t)$ is still
isomorphic to $B_3$. However for $t = 0$ we get something
interesting. We claim that
\be B_0\ :=\ \pi_{\tilde{\cal B}}^{-1}(0) \ = \ E\,\cup  \tilde
B.\label{eq:b0}
\ee
Furthermore, these two components are glued in the following way. We can think
of $E={\bf P}(\cO \oplus N_S)$ as the compactification of the
total space of $N_S$ by a section $S_\infty$ at infinity. In this
realization, the zero section of $N_S$ is identified with a section
$S_0$ of ${\bf P}(\cO \oplus N_S)$. Then the two components are
glued by identifying $S_\infty\subset{\bf P}(\cO \oplus N_S)$
with the original divisor $S
\subset B_3\simeq\tilde B$.

Let's check (\ref{eq:b0}) and the assertions below it by
writing down the defining equations for $B_0$ explicitly.
The divisor $S$ is the zero locus of a section $s$ of $\cO_{B_3}(S)$.
Thus the locus $S\times\{0\}$ that we are blowing up is given by $s = t =
0$.\footnote{Here we have chosen a global construction of the blowup.
Alternatively, one can pick local equations $f_\alpha=0$ for $S$ in open sets
$U_\alpha$ and construct the blowup locally, then glue the patches of the
blowup according to how the $f_\alpha$ change between patches.}
The blowup can be constructed as a hypersurface in the fourfold
${\bf P}(\cO_{B_3}\oplus\cO_{B_3}(S))$, a ${\bf P}^1$-bundle over $B_3$.
To partially coordinatize this fourfold,
we introduce two additional variables $u$ and $v$, with $(u,v)$
taking values in $\cO_B\oplus\cO_B(S)$, and then we identify
$(u,v)\sim(\lambda u,\lambda v)$ for any $\lambda\in{\bf C}^*$.  The
hypersurface $\tilde{\cal B}\subset{\bf P}(\cO_{B_3}\oplus\cO_{B_3}(S))$
is given by the standard equation of a blowup
\be
su - tv \ = \ 0.
\label{eq:blowup}
\ee

When $s$ and $t$ are not both zero, we find that $u$ and $v$ are
uniquely determined up to the action of ${\bf C}^*$. When $s=t=0$,
we get an extra ${\bf P}^1$ parametrized by $(u,v)$. So for each
point on $S\times \{ 0\}$, we grow an extra ${\bf P}^1$, showing
that the exceptional divisor $E$ is a ${\bf P}^1$-bundle over $S$. After
restricting
to the exceptional divisor $E$, i.e.\ setting $s=t=0$ in
(\ref{eq:blowup}), we see that $(u,v)$ take values in
${\bf P}(\cO \oplus N_S)$, since $\cO_B(S)$ restricts to $N_S$ on
$S$. This completes the demonstration that $E\simeq {\bf P}(\cO \oplus N_S)$.
The zero section $S_0$ is given by $v=0$, and $S_\infty$ is
given by $u=0$.

The total space $\tilde {\cal B}$ is smooth, but the fibers of
$\pi_{\tilde {\cal B}}$ are not. For $t \not = 0$, we get a
variety isomorphic to $B_3$ as noted above.
But for $t=0$, we get two components,
as the equation (\ref{eq:blowup}) becomes $su=0$. Away from the
zeroes of $s$, we get $u=0$, hence a copy of $B_3$ (with $S$ excised). Away
from the zeroes of $u$, we get $s=0$, hence the exceptional divisor (minus
$S_\infty$). So altogether, for $t=0$ we get a normal crossing
variety with two components.

Let's now take a closer look at the intersection $E\cap\tilde{B}$ of the two
components.  As we have seen above, $B_0=E\cup\tilde{B}$ is the subvariety
of $\tilde{\cal B}$ defined by $t=su=0$.  The two components of $B_0$ are
therefore defined by $s=t=0$ and $s=u=0$.  Since the former are the equations
of $S\times\{0\}\subset{\cal B}$, they define the exceptional divisor $E$
after pulling back to $\tilde{\cal B}$.  Since $B_3\times\{0\}$ is defined
by $t=0$, its proper transform $\tilde{B}$ is given by the remaining
component, $s=u=0$.  Finally,
the components intersect in the curve
$E\cap \tilde{B}$ defined by $s=t=u=0$.  Viewed as a curve inside $E$ it
is defined by $u=0$ hence is identified with $S_\infty$.  Viewed as a curve
inside $\tilde{B}$ it is defined by $s=0$, hence is identified with $S$.

This completes the justification of all of our claims about the geometry
of $B_0$.

Note that the normal cone degeneration itself is not a complex
structure deformation in the usual sense. All the fibers for $t
\not = 0$ are isomorphic, and at $t=0$ the fiber $B_0$ `jumps.'
There is no monodromy around $t=0$. However, nontrivial monodromy will
occur once we put a
suitable elliptic fibration over $\tilde{\cal B}$.

\newsubsection{Degenerating the elliptic fibration}

\subseclabel{LocalDegeneration}

Our next task is to identify a degeneration for the full
Calabi-Yau, including the elliptic fibration. We will do this by
extending the elliptic fibration over the one-parameter family
considered in the previous subsection, obtained from deformation
to the normal cone of a divisor $S \subset B$. This will give us a
family ${\cal Y}$, with fibers $Y_t$, such that
\be Y_0 \ = \ W_1 \cup_Z W_2 \label{eq:y0}\ee
where $W_1$ is an elliptic fibration over $E$ and $W_2$ is an
elliptic fibration over $\tilde B \simeq B_3$. Furthermore, $W_1$
will have the structure of a $dP_9$-fibration over $S$.

Let $B_3$ be a smooth three-dimensional $F$-theory base.  We can describe an
elliptically fibered Calabi-Yau $Y\to
B_3$ in generalized Weierstrass  form.  Given sections $a_i\in H^0(B_3,-i
K_{B_3})$, we can write the equation of an elliptic fibration over $B_3$
in the form
\be\label{TateForm} y^2+a_1xy+a_3y\ =\ x^3+a_2x^2+a_4x+a_6. \ee
If we now replace the $a_i$ by $t$-dependent sections $a_i(t)$, we
will get a family
\be\label{TateFormfamily} y^2+a_1(t)xy+a_3(t)y\ =\ x^3+a_2(t)x^2+a_4(t)x+
a_6(t). \ee
of elliptically fibered Calabi-Yau fourfolds
parametrized by $t$, which we can also view as an elliptically fibered
Calabi-Yau fivefold over ${\cal B}$.
We now want to understand what happens when we
pass to the normal cone degeneration by blowing up $S\times\{0\}$.  A
modification is needed since the blowup spoils the Calabi-Yau condition.

The modification is simple enough: we merely have to ensure that the $a_i(t)$
remain
sections of $H^0(\tilde{\cal B},-iK_{\tilde{\cal B}})$ after pullback to
$\tilde{\cal B}$ in order to preserve
the Calabi-Yau condition.  This would make ${\cal Y}$ a Calabi-Yau fivefold,
elliptically fibered over the fourfold $\tilde{\cal B}$.
Composing with the projection
$\pi_{\tilde{\cal B}}:\tilde{\cal B}\to {\bf C}$, we see that via
$\pi_{\cal Y}:{\cal Y}\to {\bf C},$ ${\cal Y}$ can be
viewed as a family of threefolds $Y_t$ parametrized by $t\in{\bf C}$.
Each $Y_t$, including $Y_0$, is Calabi-Yau by the adjunction formula and the
Calabi-Yau property of ${\cal Y}$, as
the divisor class of $Y_t$ is trivial since $Y_t$ is the pullback of a point
$t\in{\bf C}$ by the projection $\pi_{\cal Y}$.

The formula for the canonical bundle of the blowup gives
\be K_{\tilde{\cal B}}=\rho^*\left(K_{\cal B}\right)+E.\label{eq:blowupk}\ee
from which it follows immediately that
\be -iK_{\tilde{\cal B}}=\rho^*\left(-iK_{\cal B}\right)-iE.
\label{eq:cycondition}\ee
Then we read off from (\ref{eq:cycondition}) the desired Calabi-Yau condition
as the requirement that each $a_i(t)$ vanishes to order at least $i$ along
$S\times\{0\}$.  Given such an $a_i(t)$, its pullback to $\tilde{\cal B}$
vanishes along $E$ with multiplicity at least $i$.  Letting
$e\in H^0({\cal O}_{\tilde{\cal B}}(E))$ be a section vanishing along $E$, we see that
we have holomorphic sections
\be \tilde{a}_i(t):=\frac{a_i(t)}{e^i}\in
H^0(\rho^*\left(-iK_{\cal B}\right)-iE)=
H^0\left(-iK_{\tilde{\cal B}}\right)\ee
which can be used to construct the desired elliptically fibered ${\cal Y}\to
\tilde{\cal B}$ in the usual way.

This gives the desired family of F-theory models.  For $t\ne0$, $Y_t$ is
elliptically fibered over $B_3$, but $Y_0$ is given by (\ref{eq:y0})
as a union of elliptic fibrations $W_1$ and $W_2$, fibered over the
respective components
$E$ and $\tilde{B}\simeq B_3$ of $B_0$.

The intersection $Z$ of $Y_1$ and $Y_2$ is elliptically fibered over the surface
$S=E\cap \tilde{B}$.  In fact, $Z$ is an elliptically fibered Calabi-Yau
threefold.  We just have to check that the $\tilde{a}_i(t)$ restrict to
sections of $-iK_S$ on $S$, and for that it suffices to show that
$K_{\tilde{\cal B}}$ restricts to $K_S$ on $S$.  But that follows from the
adjunction formula: $S=E\cap \tilde{B}$ implies that
\be K_S=\left(K_{\tilde{\cal B}}+E+\tilde{B}\right)|_S,\ee
while $E+\tilde{B}=B_0$ is a fiber, which we have already noted is
trivial as a divisor class.

\smallskip\noindent
The restriction of $K_{\tilde{\cal B}}$ to $B_0$ is in fact the
{\em dualizing sheaf\/} of $B_0$, denoted $\omega_{B_0}$,
which plays the role of the canonical
bundle for $B_0$.  This observation will be useful in Section~\ref{Example}
when we glue local models.

We can describe $\omega_{B_0}$ in terms of its restriction to the components
$\tilde{B}$ and $E$ of $B_0$.  From (\ref{eq:blowupk}) we read off that
\be \omega_{B_0}|_{\tilde{B}}=K_{\tilde{B}}+S. \ee
We also have by adjunction that $K_E=(K_{\tilde{B}}+E)|_E=\omega_{B_0}|_E-
S_\infty$, hence
\be \omega_{B_0}|_E=K_E+S_\infty. \ee
In other words, sections of $\omega_{B_0}$ on each component can be identified
with sections of the canonical bundle with first order poles on the
intersection $S$.

The restriction of $K_{\tilde{B}}$ to $B_t$ for $t\ne0$ is just the ordinary
canonical bundle of $B_t$.  In fact, $K_{\tilde{B}}$ is the
{\em relative dualizing sheaf\/} $\omega_\pi$ of the family $\pi_{\tilde{B}}$.

\smallskip
We will now show that $W_1$ has the structure of a $\mathrm{dP}_9$
fibration over $S$.
Recall the a $\mathrm{dP}_9$ arises as an elliptic fibration over ${\bf P}^1$
constructed from (\ref{TateForm}) with $a_i\in H^0({\bf P}^1,{\cal O}_{{\bf P}^1}
(i))$.  So we just have to show that $-K_{\tilde{\cal B}}$ restricts to
${\cal O}_{{\bf P}^1}(1)$ on the fibers of the projective bundle
$E={\bf P}({\cal O}\oplus N_S)$.  But this follows immediately from
(\ref{eq:blowupk}) as $\rho^*(K_{\cal B})$ has trivial restriction to the
fibers while ${\cal O}_{\tilde{\cal B}}(E)$ restricts to ${\cal O}_{{\bf P}^1}(-1)$.

Physically the reason we are interested in this degeneration is because
we may have singular fibers corresponding to an enhanced gauge symmetry
along $S$, and we want to capture this geometry in a local model.
So we now specialize (\ref{TateForm}) to one of the Tate forms
\cite{Tate,Bershadsky:1996nh} which
imply that we have enhanced gauge symmetry along $S$.  We investigate
when we can follow this singularity through the degeneration to the normal cone
and  preserve the singularity type when we get to $B_0$, while ``pushing
the singularity away from $B_3$''
by putting the singularity at $S_0\subset E$, which is disjoint from
$\tilde{B}\simeq B_3$.
For simplicity, we stick to simply laced cases where we do not need to
worry about monodromy issues.  So we simply need to control the order of
vanishing of the $a_i$ along $S$ and of the $\tilde{a}_i(t)$ along $S_0$.
Let's say that the desired order of vanishing of $a_i$ along $S$ is $n_i$ for
the appropriate gauge group. We need to assume the hypothesis
that $n_i\le i$. The ADE types
of such singularities correspond to subgroups of $E_8$.

\smallskip\noindent
{\bf Claim}: Suppose that the $a_i(t)$ satisfy the Calabi-Yau
condition that it vanishes to order at least $i$ on $S
\times \{0\}$, and furthermore that the $a_i(t)$ vanish to at least
order $n_i$ on all
of $S\times {\bf C}$. Then after the blowup which realizes the normal
cone degeneration, the order of vanishing of $\tilde{a}_i(t)$
along $S_0\subset E\subset B_0$ is also at least $n_i$.

\smallskip\noindent
Proof: By our assumptions, we can write
$a_i(t)=s^{n_i}f_i(t)$ where $f_i(t)$ is
holomorphic.  Then $f_i(t)$ vanishes to order at least $i-n_i$ along $E$ after
pulling back to $\tilde{\cal B}$.
Since we only need the order of vanishing of $\tilde{a}_i(t)$
along $S_0$,
we can compute in local coordinates near $S_0$.
We have seen that $S_0\subset E$ is defined by $v=0$, hence $v/u=s/t$
is an affine $N_S$-valued
fiber coordinate on the projective bundle $E={\bf P}({\cal O}
\oplus N_S)$, vanishing along $S_0$.  In this coordinate patch, $E$ is defined
by $t=0$.  It follow that in this patch $f_i(t)$ is divisible by $t^{i-n_i}$
and also that we may write
$\tilde{a}_i(t)\ =\ a_i(t)/t^i$.
Then we compute that
\be \tilde{a}_i(t)=\frac{a_i(t)}{t^i}=\left(\frac{s}{t}\right)^{n_i}
\left(\frac{f_i(t)}{t^{i-n_i}}\right)\ee
with $f_i(t)/t^{i-n_i}$ holomorphic.  It follows
that $\tilde{a}_i(t)$ vanishes to order at least $n_i$ along $S_0$ as
claimed.

\smallskip
{\bf Example.} We can achieve an $E_8$ gauge symmetry
by requiring the order of vanishing of $(a_1,a_2,a_3,a_4,a_6)$ to be at least
$(1,2,3,4,5)$.  If we can choose sufficiently generic sections
${a}'_i\in H^0(B,-i(K_B+D))$ for $i=1,2,3,4$, and
${a}'_6\in H^0(B,-6K_B-5D)$, then the elliptic fibration
\be
y^2+{a}'_1sy+{a}'_3s^3=x^3+{a}'_2s^2x^2+{a}'_4s^4x+
{a}'_6s^5
\ee
over $B_3$ has an $E_8$ singularity along $S$.

For $i=1,\ldots,4$, it is automatic that $a_i(t)=a_i'(t)s^i$ vanishes to
order at least $i$ along $S\times\{0\}$.  But for $i=6$, ${a}'_6(t)s^5$
need not vanish to order~6 along $s=t=0$ as required to get a Calabi-Yau.
One way to achieve the desired vanishing is to replace ${a}'_6(t)s^5$
with ${a}'_6(t)s^5t$.  Thus the desired degeneration is given by
\be
y^2+{a}'_1(t)sy+{a}'_3(t)s^3=x^3+{a}'_2(t)s^2x^2+{a}'_4(t)s^4x+
{a}'_6(t)s^5t.
\ee
To obtain the elliptic fibration $W_2$ over $\tilde{B}$, we divide $a_i(t)$
by $s^i$ and put $t/s=0$
to obtain
\be
y^2+{a}'_1y+{a}'_3=x^3+{a}'_2x^2+{a}'_4x,
\ee
where we have put $a_i'=a_i'(0)$.

To obtain the elliptic fibration $W_1$ over $E$, we instead
divide each $a_i'(t)$ by
$t^i$, introduce the local equation $\tilde{s}=s/t$ which vanishes along
$S_0$, and set $t=0$.  That gives
\be
y^2+{a}'_1|_S\tilde{s}y+{a}'_3|_S\tilde{s}^3=x^3+{a}'_2|_S\tilde{s}^2x^2+{a}'_4|_S\tilde{s}^4x+{a}'_6|_S\tilde{s}^5, \label{eq:ellipticw1}
\ee
which has the required $E_8$ along $S_0$.

Another example is discussed in section \ref{Example}.

\newsubsection{Caveats}

Our construction made use of the Tate form of the Weierstrass
fibration.  As was recently investigated in \cite{Katz:2011qp}, it
is not clear if every elliptic fibration with, say,
$SU(5)$ singularities along a divisor $S$ can be put in this form.
Although no
counterexample was found, we see no a priori reason why it should be
the case that every elliptic fibration with $SU(5)$ singularities along
$S$ can be put in Tate form.

For the purpose of constructing global models along the lines of
\cite{Donagi:2009ra}, the more relevant question however is not
whether one can put the elliptic fibration in Tate form, but
whether there exists a degeneration to a local model. If the
elliptic fibration can be put in Tate form, then as we saw above
it is fairly straightforward to construct such a degeneration. It
is not clear to us how to construct such a degeneration directly
from the Weierstrass form, as the conditions for having suitable
singularities along $S$ are rather non-linear in this formulation.
But this does not mean that such a degeneration does not exist.
Indeed, in the next section we will see another approach to
constructing global Calabi-Yau manifolds, where the existence of a
degeneration is built in from the start. This method makes no
reference to the Tate form, only to the Weierstrass form.

On the other hand, it clearly should not be true that every single
elliptic fibration with $SU(5)$ singularities along $S$ admits such
a degeneration. Let us suppose that along a sublocus of $S$, the
$SU(5)$ singularity degenerates to a type of singularity that cannot
be fit in $E_8$, like say $SU(100)$. Then plausibly the degeneration
to a local $dP_9$ model should not exist, or should develop very bad
singularities. As a rule of thumb, it seems reasonable to expect
that the degeneration to the local $dP_9$ model exists if the
singularities can be fit in $E_8$, as we saw explicitly in our
construction above through the condition $n_i \leq i$. Fortunately
this appears to be sufficient for phenomenological applications. For
example for $SU(5)_{GUT}$ models, generically the worst we expect is
singularities of type $E_6$, $SO(12)$ and $SU(7)$ on $S$, all of
which can be fit in $E_8$.\footnote{We are referring here to the
singularity type of the fiber over a point in $S$, not the
singularity type of the total space. It is the former that is
relevant for the spectral cover/gauge theory approach.} Note that
$E_6$ cannot be fit in any $SO(n)$, so we expect the Sen limit to be
problematic, as was indeed found in \cite{Donagi:2009ra}.

At any rate, the global-to-local degeneration gives a clearer
justification for the use of Higgs bundle and spectral cover
technology, since a $dP_9$-fibration with ${\sf G}$-flux is
equivalent to spectral data. Similarly it gives a better
justification for the `heterotic' approach to instanton effects in
$F$-theory \cite{Donagi:2010pd}. In the remainder, we would like
to place this in the general context of weak coupling limits and
the cutting-and-gluing approach.

\newpage

\newsection{Gluing a Calabi-Yau from local pieces}

\seclabel{CYGluing}

We have motivated the issue of degeneration on phenomenological
grounds, and from the point of view of string dualities and weak
coupling limits. However there is another good motivation, based on
an analogy with topological field theories.

\newsubsection{Some aspects of topological field theories}

Let us give some motivation by first considering topological field
theories \cite{AtiyahTop}. We will not give a systematic
exposition, but rather discuss a few general features that will be
useful below.

A topological field theory on a closed manifold $M$ is a quantum
field theory in which the correlation functions are independent of
the background metric. Our main interest here is in the gluing
axiom. If the space $M$ has a boundary $\del M = \Sigma$, then the
field theory produces a state in a Hilbert space ${\cal H}_\Sigma$
given by quantizing the fields on the boundary. Now we can take
two manifolds $M_1$ and $M_2$ with boundary with boundary
$\Sigma$, and glue them together. The gluing axiom states how the
the correlation functions are obtained from $M_1$ and $M_2$. For
the partition function, this takes the form
\be\label{GlueAxiom} Z_M \ = \  \sum_{v_i\in {\cal H}_\Sigma}
Z_{M_1}(v_i) Z_{M_2}(v^i) \ee
where $v_i$ is a basis of ${\cal H}_\Sigma$ and $v^i$ is the dual
basis of ${\cal H}^*_\Sigma$.

The partition function of a topological field theory on a manifold
$M$ can thus be computed by cutting the manifold into pieces and
applying the rule above. For each piece, the field theory
determines a distinguished state in the tensor product of Hilbert
spaces associated to the boundaries. If the pieces are simple
enough, then we may be able to write down this state explicitly.
Then we glue the pieces back together to obtain the partition
function on $M$. For beautiful examples of this paradigm, see eg.
\cite{Witten:1988hf} or \cite{Witten:1991we}.

For the discussion below, it will also be useful to recall some
aspects of Chern-Simons theory on a real $3d$ manifold $M_3$,
which we take to be closed for now. The  way the
Chern-Simons action is usually defined in the literature is
as follows: we pick a real $4d$ manifold
$M_4$ such that $M_3 = \del M_4$, and we extend the gauge field
$A$ over $M_4$. Then the action is defined as
\be S_{CS}(A) \ = \ {k\over 8\pi^2} \int_{M_4} {\rm Tr}(F_A\wedge
F_A). \ee
One may show that this does not depend on extension that one picks,
up to an integral ambiguity. Namely if one chooses a different
extension, then by gluing the two choices into a closed manifold and
evaluating the above action, we see that the difference between the
two choices must be an integer, so that $e^{iS_{CS}}$ is well defined.

The extension is not always possible, and a more complete answer has
been given in \cite{Dijkgraaf:1989pz}. In fact it will be useful to
consider the definition of the Chern-Simons-type invariants on a general
manifold $X$, with dimension possibly greater than three. A bundle
on $X$ can be thought of as an equivalence class of maps from
$\gamma:X \to BG$, where $BG$ is the classifying space of the gauge
group $G$. The bundle on $X$ is then identified with the pull-back
of the universal bundle $EG$ over $BG$. We will take $G$ to be
connected and simply connected, as this simplifies things a little
bit.

Since characteristic classes behave naturally under pull-back, we
can now consider the analogous problem for the image of the
three-cycle $M$ under the map $\gamma$. Namely we ask for a bounding
four-chain for $\gamma(M)$. For $G$ connected and simply connected,
we have $H_3(BG,{\bf Z})=0$, so the homology class $\gamma_*[M] = 0$
and such a bounding four-chain exists, which then comes equipped with a
$G$-bundle.  We then extend the connection and integrate
${1\over 8\pi^2}{\rm Tr}(F\wedge F)$ as before. For higher dimensional $X$,
the Chern-Simons action is not canonically defined if the first Pontryagin
class does not vanish on $X$, and we subtract a contribution
${1\over 8\pi^2}{\rm Tr}(F_0\wedge F_0)$ of a reference connection $A_0$.

For general $G$, we have to worry about issues resulting from torsion in
the cohomology of $BG$, so we can only bound an integer multiple $n\gamma$
of $\gamma$.  We can integrate over a bounding chain as before and then divide
by $n$, but the result will only be well-defined up to $1/n$ times an
integer.  To properly
define the theory, have to fix a cohomology
class in $H^4(BG,{\bf Z})$ which can be used to fix the ambiguity.
When $G$ is connected and simply connected, we have
$H^4(BG,{\bf Z})={\bf Z}$ and we can just use the
first Pontryagin class  of the
universal bundle.

The choice of extension is not unique, and by changing the choices,
we again get a natural discrete ambiguity. When $X$ is a
three-manifold, this gives an integer ambiguity. For more general
$X$, we get such an integer ambiguity for every three-cycle, in
other words the Chern-Simons form is defined only modulo the lattice
$H_3(X,{\bf Z})$.

When $X$ has a boundary, then we can proceed similarly. We glue on
another manifold with boundary to get a closed manifold, extend the
gauge field, and then apply the discussion above. The result depends
on a choice of extension, which introduces an ambiguity, and is not
quite gauge invariant. Under a gauge transformation which is
non-trivial on the boundary, we have
\be S(A) \ \to \ S(A) + {k\over 8\pi^2}\int_{\del M} {\rm Tr}( A
\wedge dg g^{-1}) -{k\over 24\pi^2} \int_M{\rm Tr}(g^{-1}dg)^3 \ee
where we can see the WZW term emerging. For simplicity, we assumed
here that the bundle is trivial. In fact, Chern-Simons theory on
$D \times {\bf R}$ (with $D$ a disk) is equivalent to a chiral
sector of a WZW model on the boundary $S^1 \times {\bf R}$
\cite{Moore:1989yh}.

If the boundary is empty, then we see from the above expression
that the Chern-Simons action shifts under a topological quantity,
the degree of the map $g:M \to G$. This gives another
interpretation of the ambiguities: they arise from large gauge
transformations that are not continuously connected to the
identity.

\newsubsection{Analogy between topological and holomorphic field theories}

\subseclabel{HolomCS}

In holomorphic field theories, there is obviously some dependence
on the metric on $M$. Furthermore we cannot simply cut up a
manifold without breaking holomorphy, so the conventional notion
of a boundary and the gluing axiom is not so useful. Nevertheless
there is an analogue of the cutting and gluing procedure in
topological field theories which is compatible with holomorphy. It
is given by the degeneration of the complex manifold to a normal
crossing variety, like the degeneration we discussed in the last
section.

For physicists, this is familiar for example from the gluing
construction of conformal blocks in $2d$ CFT. We can degenerate the
Riemann surface into pieces joined by a long neck. The neck is
described algebraically by the equation
\be\label{ZDeg} xy \ = \ q \ee
In the limit $q\to 0$, the Riemann surface degenerates to a union
$\Sigma_1 \cup_p \Sigma_2$ glued along a puncture $p$. Instead of
(\ref{GlueAxiom}) above, for small $q$ we can construct conformal
blocks using a $q$-expansion of the form
\be Z_\Sigma(R,q) = \sum_{e_i\in R} Z_{\Sigma_1}(e_i) q^{w(e_i)}
Z_{\Sigma_2}(e^i). \ee
We want to do something similar with higher dimensional varieties.
A suitable higher dimensional analogue of (\ref{ZDeg}) is the
normal crossing singularity.

A further intriguing aspect of such a picture is the interplay
between the theory in the bulk and the theory on the boundary. In
fact even though the theories we are dealing with are not
topological, in a number of known examples we can express the
theory at small coupling entirely on the boundary. For $F$-theory
the previously known degeneration limits are the $SO(32)$ and
$E_8\times E_8$ heterotic limits. In these limits, the Calabi-Yau
splits into two pieces, and the heterotic theories may be thought
of as living on the normal crossing divisor
\cite{Morrison:1996pp,Aspinwall:1997ye,Clingher:2003ui}, which we
interpret as the boundary. In \cite{Donagi:2010pd,CDW}, the Sen
limit is discussed from the same point of view and gives a further
example: the boundary is the IIB space-time. The superpotential
can be computed either on the fourfold or in the boundary theory.

We would like to argue that this should hold more generally.
Instead of the algebraic arguments that have been used
for the heterotic and IIB limits, we want to discuss an argument
that is more in the spirit of topological field theory. The
holomorphic analogue of real and orientable $n$-dimensional
manifold with boundary is an $n$-complex dimensional log
Calabi-Yau manifold. The holomorphic analogue of the orientation
of a real manifold is holomorphic $(n,0)$-form on the log
Calabi-Yau, and the holomorphic analogue of the boundary is the
polar divisor of the $(n,0)$-form, i.e. the log divisor
\cite{Donaldson:1996kp,Khesin:2000ng}. The statement below was
already given fifteen years ago \cite{ThomasGaugeTheory}. We will
give a slight variation of the original argument.

The analogue of Chern-Simons theory on $M_3$ is holomorphic
Chern-Simons theory on a Calabi-Yau three-fold $X_3$. We will
proceed as above for conventional Chern-Simons theory: we try to
extend our gauge fields on $X_3$ to a log Calabi-Yau fourfold
$(X_4,X_3)$, whose boundary is $X_3$. Then we try to define the
holomorphic Chern-Simons action as
\be\label{HCSExtension} S_{CS}(A) \ = \ {1\over 8\pi^2}\int_{X_4}
\Omega^{4,0} \wedge {\rm Tr}(F_A\wedge F_A) \ee
This action would normally seems to depend on the extension one
picks. The choice of extension may not be unique, and as before
this gives rise to a natural ambiguity. Furthermore, the holomorphic
Chern-Simons action is in general not canonically defined, but
depends on a choice of base point. We can eliminate the dependence
on a reference connection only when
$p_1|_{X_3}$ vanishes. If $p_1$ is non-vanishing, then we
cannot assign the same value of the Chern-Simons action to every
three-cycle in the same homology class. (This is very similar to
the Picard group: if $c_1(L)$ does not vanish, then we do not get
a canonical map from the line bundle $L$ to the Jacobian. Instead
we get a torsor structure, an the combined data defines a differential 
character or Deligne cohomology class). However when tadpole cancellation is
satisfied, eg. $p_1(V_1) + p_1(V_2) - p_1(TX_3)=0$ in the
heterotic string, then the sum of the Chern-Simons contributions
is defined. If one needs $5$-branes wrapped on curves to satisfy
tadpole cancellation, then their contribution to the superpotential
also has to be included.

We now use the fact that the logarithmic cohomology $H^k_{\rm
log}(X_4)$ for our pair $(X_4,X_3)$ is isomorphic to $H^k(U)$,
where $U = X_4 \backslash
X_3$. Further we have $H^{2n-k}(X_4,X_3)\cong H^{2n-k}_c(U)$,
where $H^*_c$ denotes compactly supported cohomology. It follows
that we have a Poincar\'e duality map $H^k_{\rm log}(X_4)\to
H_{2n-k}(X_4,X_3)$. From the short exact sequence
$0\to \Omega^p_{X_4}\to\Omega^p_{X_4}(\log X_3)\to
\Omega^{p-1}_{X_3} \to 0$ we get the following long exact
sequence for $H^k_{\rm log}(X_4)$:
\be \ldots\ \to\ H^{k-2}(X_3)\ \to\ H^k(X_4)\ \to\ H^k_{\rm
log}(X_4)\ \mathop{\to}^{\rm res}\ H^{k-1}(X_3)\ \to\ \ldots \ee
This is precisely the Poincar\'e dual to the long exact sequence for
relative homology
\be
 \ldots \to H_{2n-k}(X_3) \to H_{2n-k}(X_4) \to H_{2n-k}(X_4,X_3)
\mathop{\to}^{\del} H_{2n-k-1}(X_3) \to \ldots \ee
Now let us take a class $\alpha = p_1(A)-p_1(A_0)$ in $H^4(X_4)$
which restricts to a trivial class in $H^4(X_3)$. From the long
exact sequence
\be \ldots\ \to\ H^4(X_4,X_3)\ \to\ H^4(X_4)\ \to\ H^4(X_3)\ \to\
\ldots \ee
we see that $\alpha$ lifts to $H^4(X_4,X_3)\cong H^4_c(U)$. So now
we compute
\ba \int_{X_4}\Omega \wedge (p_1(A)-p_1(A_0)) &=& \int_D
p_1(A)-p_1(A_0) \eol &=& \int_{\del D}\omega_{CS}(A) -
\omega_{CS}(A_0) \eol &=& 2\pi i \int_{X_3}{\rm res}(\Omega)
\wedge (\omega_{CS}(A) - \omega_{CS}(A_0)) \ea
where $D \in H_4(X_4,X_3)$ is the Poincar\'e dual of $\Omega$.
This is what we wanted to show.

Note that on a log Calabi-Yau, we can rewrite
$\int_{X_4}\Omega^{4,0}\wedge \alpha$ as an expression on $X_3$
for any $\alpha\in H^4(X_4)$, whether or not the restriction to
$H^4(X_3)$ is trivial. To see this, we note that the $(0,4)$ part
of $\alpha$ is Dolbeault exact, so with $\alpha^{0,4} = \delb
\omega$ we can write
\be \int_{U} \Omega^{4,0} \wedge \delb \omega \ = \ 2\pi i
\int_{X_3}\Omega^{3,0} \wedge \omega|_{X_3} \ee
where we used the Stokes and Cauchy theorems. However to relate
this to the Chern-Simons action, we need the assumption on the
first Pontryagin class.

In general it may not be possible to extend the gauge field $A$ in
this way, but we can still fall back to the definition of the
Chern-Simons form in \cite{Dijkgraaf:1989pz}, which we reviewed
above. However we see that there's a further issue. The
superpotential in $4d$ $N=1$ supergravity is a section of a line
bundle ${\cal L}$ over the moduli space, so there is no room for
periodic identifications. Also, the ambiguities of the holomorphic
Chern-Simons action correspond to the periods of $\Omega^{3,0}$,
which are usually dense in ${\bf C}$ (for example if the rank of
$H_3(X_3,{\bf Z})$ is at least three). This is not a problem if
one is only interested in the derivatives, but it seems
problematic whenever we are interested in the value of the action.

This issue reappears in superpotentials for brane configurations
more generally. For example the superpotential for a brane wrapped
on a two-cycle $C$ in a Calabi-Yau three-fold is given by
\cite{Witten:1997ep}
\be\label{CurveSuperW} W \ = \ \int_\Gamma \Omega^{3,0} \ee
where $\del \Gamma = C - C_0$ and $C_0$ is a holomorphic reference
curve in the same homology class. This superpotential is actually
closely related to the holomorphic Chern-Simons action
\cite{Donaldson:1996kp,ThomasGaugeTheory,TyurinAbel,ClemensDef}.
For holomorphic bundles, the Chern-Simons action is independent of the
choice of hermitian metric and can be written
as (\ref{CurveSuperW}) where $C-C_0$ is the Chern character
${\rm ch}_2$, thought of as a Chow class \cite{TyurinAbel}. As noted in
\cite{Morrison:2007bm}, the description using ${\rm ch}_2$ is
then naturally extended to coherent sheaves and even the derived
category. By changing the choice of $\Gamma$, again we have an ambiguity by
$H_3(X,{\bf Z})$, whose image under pairing with $\Omega^{3,0}$ is
generally dense.

Nevertheless in string compactification this problem of periodic
identifications is never an
issue, as there are additional terms in the superpotential.%
\footnote{We are grateful to E.Witten for illuminating
comments in this regard.} %
This is another instance of the principle that the individual
terms may be problematic, and only the sum of the terms has to be
well-defined. For example in the heterotic string we have $W \sim
\int_X\Omega^3 \wedge (H + \omega_{CS})$ with $dH + d\omega_{CS} =
0$. Since $dH$ is not identically zero, $H$ does not live in
$H^3(X,{\bf Z})$ but in a torsor, a principal fiber bundle with
structure group $H^3(X,{\bf Z})$. However $H + \omega_{CS}$ does
define a class in $H^3(X,{\bf Z})$, and $W$ is well-defined up to
the usual ambiguity in picking $\Omega^{3,0}$. A similar
phenomenon occurs in type IIB.

We believe that this perspective illuminates the role of
Calabi-Yau fourfolds in the hierarchy of holomorphic field
theories. The above expression (\ref{HCSExtension}) is of course
very reminiscent of the flux superpotential that one encounters in
$F$-theory. Indeed, it seems very natural to think about the flux
superpotential in this way. Given an $F$-theory compactification
with $G$-flux in a degeneration limit, let us consider the part of
the flux which comes from $W_1$ or $W_2$ separately. We will
briefly mention the other possible fluxes later.

Up to a shift in the quantization law, we may always express the
cohomology class of the $G$-flux as ${\sf G}/{2\pi} = p_1(V)$ for
some bundle $V$ on $Y$. According to \cite{Witten:1996md} the
shift is given by $- \half p_1(TX_4)$ on a compact real
eight-dimensional manifold. In our context we have instead a log
Calabi-Yau manifold $(X_4,X_3)$. We propose that the tangent
bundle $TX_4$ should be replaced by the logarithmic tangent bundle
$\Theta \equiv TX_4(-\log X_3)$, which is the subbundle of $TX_4$
generated in coordinates by $\partial/\partial z_1,\
\partial/\partial z_2,\
\partial/\partial z_3$, and $z_4\partial/\partial z_4$, where
$z_4=0$ is a local equation  for $X_3\subset X_4$.  If $X_4$ is a
component of a normal crossings variety $Y_0=X_4\cup X_4'$ whose
components intersect along $Z=X_3$, then $T{X_4}(-\log X_3)$ is
the restriction to $X_4$ of the logarithmic tangent bundle
$T{Y_0}(-\log Z)$ introduced in section \ref{SemiStable}.

Thus we can write the superpotential as
\be W \ = \ {1\over 2\pi}\int_{X_4}\Omega^{4,0} \wedge {\sf G} \ =
\ \int_{X_4}\Omega^{4,0} \wedge \left[p_1(V)- \half
p_1(\Theta)\right] \ee
Here we are implicitly assuming that ${\sf G}|_{X_3}$ vanishes in
$H^4(X_3)$, otherwise we should subtract the contribution of a
reference connection as discussed above. Then as before we get:
\ba\label{Transgression} W& =&  \int_{D}  \left[p_1(V)- \half
p_1(\Theta)\right]  \eol &= & 2\pi i\int_{X_3} \Omega^{3,0} \wedge
\left[\omega_3(V|_{X_3})- \half \omega_3(TX_3)\right] \ea
with $\omega_3$ denoting the Chern-Simons form. Here we have used
that $p_1(\Theta)|_{X_3}=p_1(TX_3)$. This follows from the exact
sequence
\be 0\ \to\ TX_3\ \to\ TX_4(-\log X_3)|_{X_3}\ \to\  {\cal
O}_{X_3}\ \to\  0\ee
which is easily checked using local coordinates.

For the local $dP_9$-fibration discussed in the previous section,
this ought to reproduce the holomorphic Chern-Simons theory on the
boundary $Z$ with a single $E_8$ gauge group, if we restrict to
allowed $F$-theory fluxes.\footnote{It is natural to conjecture an
analogous holographic relation for a $G_2$-manifold with
superpotential $W = \int (\omega_3(A) + i \Phi)\wedge
d(\omega_3(A) + i\Phi)$ and holomorphic Chern-Simons theory on its
Calabi-Yau threefold boundary.} This has previously been noted in
\cite{Jockers:2009ti}. We will show in section \ref{WAsymptotics}
using the cylinder mapping for $dP_9$-fibrations that at least we
get the same critical set, corresponding to holomorphic $E_8$
bundles. But the point is really that this is a rather general
property of periods on a log Calabi-Yau, and so quite generally we
get such a Chern-Simons theory living on the boundary for each
local piece of our global model.

Besides the fluxes above, there are additional ${\sf G}$-fluxes on
$W_1 \cup_Z W_2$. They come either from $H^3(Z,{\bf Z})$ using the
boundary map of the Mayer-Vietoris sequence, or from ${\sf
G}$-fluxes that have a pole along $Z$. This gives rise to extra
terms in the superpotential on the boundary of the form
$\int_{X_3} \Omega^{3,0} \wedge H$. We can understand this
intuitively by dualizing the ${\sf G}$-fluxes to homology classes,
so that we can interpret the superpotential as a period integral.
The ${\sf G}$-fluxes considered above are cohomology classes on a
local piece, so they dualize to a homology class on that local
piece. The superpotential $\int_{X_3} \Omega^{3,0} \wedge H$
instead comes from periods of $\Omega^{4,0}$ over vanishing
cycles, i.e. homology classes which disappear in the $t\to 0$
limit. In addition, there are typically also four-cycles which are
not closed on the local pieces individually, and have to `pass
through the neck.' This is easily seen using the Mayer-Vietoris
sequence. Periods over such four-cycles diverge with $\log(t)$ and
give perturbative corrections to the superpotential on the
boundary. Finally, all these periods receive corrections which are
analytic in $t$, which we should interpret as instanton
corrections on the boundary. We will discuss this in a more
precise language in section \ref{VHS}.

From here it is a natural step to conjecture that in a degeneration
limit, the full theory may be thought of as living on the boundary,
not just the holomorphic sector for which we argued above. As
evidence that this is indeed correct, we can again cite the
heterotic and IIB limits of $F$-theory.

Let us make some side remarks. One can try to generalize many
other known arguments used for topological field theories. A
particularly interesting one is the `holographic' relation between
Chern-Simons theory on $M_3$ and WZW-models on $M_2 =\del M_3$,
which should have an analogue in a relation between holomorphic
Chern-Simons theory on a log Calabi-Yau three-fold $(X_3,X_2)$ and
a $4d$ version of WZW models on $X_2 = \del X_3$. Under a gauge
transformation, we have
\be S(A) \ \to \ S(A) + {1\over 8\pi^2}\int_{\del X_3}\Omega^{2,0}
\wedge {\rm Tr}( A \wedge dg g^{-1}) -{1\over 24\pi^2} \int_{X_3}
\Omega^{3,0} \wedge {\rm Tr}(g^{-1} dg)^3  \ee
where we see a holomorphic version of the WZW term emerging. Note
that the last two terms only depend on the values of the fields at
the boundary, and we have ambiguities corresponding to periods.
This holomorphic WZW term was previously studied in
\cite{Frenkel:1996zb}.

It could also be interesting to consider Grassmann versions of
these theories, which are used in the context of the twistor
string approach to $N=4$ super Yang-Mills theory
\cite{Witten:2003nn}. The analogy with $F$-theory suggests that
instanton corrections in the twistor string can be computed with
Hodge theoretic methods by lifting to a higher dimensional space.
One can see hints of this by thinking of $D3$-branes as obtained
from compactifying $M5$-branes on an elliptic curve. In the weak
coupling limit, the elliptic curve becomes nodal and the
$D3$-brane theory should be thought of as living on the nodal
divisor. One would like a twistor version of this relation.

As a further side remark, another part of the analogy says that the
analogue of correlators of Wilson loops in Chern-Simons theory are
invariants associated to holomorphic curves in the holomorphic
Chern-Simons theory
\cite{ThomasGaugeTheory,Khesin:2000ng,FrenkelTodorov}. In the
abelian case, the holomorphic Wilson loop operator associated to a
curve $C$ is defined as
\be {\cal W}_C \ = \ \exp \int_C \Omega^{1,0} \wedge A^{0,1} \ee
where $\Omega^{1,0}$ is an orientation on $C$. In the non-abelian
case, one would presumably define it as ${\rm Det}\delb_A|_C$. In
the twistor context, it has been found that the amplitudes of $N=4$
Yang-Mills theory can be computed from such holomorphic Wilson loops
\cite{Bullimore:2011ni}. This indicates an interesting kind of
duality between instanton corrections in supertwistor space and
holomorphic Wilson loops in a dual supertwistor space. It would be
very interesting if some version of this duality exists more
generally.

At any rate, we see that from the above point of view it is very
natural to imagine constructing a Calabi-Yau manifold by gluing
together local pieces, independent of any phenomenological
considerations.

The Calabi-Yau obtained in this way will have normal crossing
singularities. For many purposes, such a singular Calabi-Yau is
practically as good as a smooth Calabi-Yau. The main difference is
that instead of ordinary differential forms, we should consider
logarithmic differential forms $\Omega^\bullet(\log D)$, which
behave nicely under a smoothing. Nevertheless one would like to know
if the normal crossing singularities can indeed be smoothed, for
otherwise one might doubt that the theory exists for finite $t$. For
applications to $F$-theory, one would also like to know if the
Calabi-Yau obtained from smoothing the normal crossing singularities
admits an elliptic fibration with a section.

\newsubsection{Smoothing criteria of Kawamata/Namikawa}

\subseclabel{KawNamSmoothing}

 Thus we now consider a Calabi-Yau
$Y_0$ obtained from gluing together a collection of local models
into a normal crossing variety. Then we would like to know if we
can deform it to a smooth global model.

According to Kawamata and Namikawa \cite{KawNam}, the normal
crossing variety can be smoothed if it admits a log structure, and
some further mild conditions.
See e.g.\ \cite{OgusLectures,LogModuli,GrossTropical} for many
results about log geometry.

Kawamata and Namikawa consider the rather general set-up where the
singular Calabi-Yau $d$-fold is of the form $\cup_{Z_j} W_i$. Then
they prove that a smoothing exists if $Y_0$ is K\"ahler, admits a
log structure, and has $H^1(Y_0) = H^{d-1}(\tilde Y_0)=0$, where
$\tilde Y_0$ denotes the normalization of $Y_0$. Locally, $Y_0$ is
isomorphic to a hypersurface $z_1 \cdots z_r = 0$ in a neighbourhood
of the origin in ${\bf C}^{d+1}$, and the deformation corresponds
to $z_1 \cdots z_r = t$ locally.

For simplicity we will only consider the case $Y_0=W_1\cup_Z W_2$.
Concretely, Kawamata and Namikawa prove that a smoothing of $W_1
\cup_Z W_2$ exists in this case under the following conditions:
\begin{enumerate}
  \item $Z \in |-K_{W_i}|$; in other words, each pair $(W_i,Z)$ is a
  log Calabi-Yau.
  \item $H^1(\cO_{W_i})=H^{d-1}(\cO_{W_i})=0$. By a Mayer-Vietoris
  argument, this gives $H^1(Y_0)=0$, and clearly $H^{d-1}(\tilde
  Y_0)=0$ since $\tilde Y_0 = W_1 \coprod W_2$.
  \item $Y_0$ should be K\"ahler. This is satisfied if there are ample divisors $H_i$ on $W_i$ such that $H_1|_Z$
  is linearly equivalent to $H_2|_Z$; then we get an ample line bundle on $Y_0$ yielding a projective
  embedding, and thus it follows that $Y_0$ is K\"ahler.
  \item $N_{Z/W_1}\otimes N_{Z/W_2} = \cO_Z$. This is called the $d$-semi-stability condition \cite{FriedmanSmoothing},
  where $d$ is the dimension of $W_i$. By proposition 1.1 of \cite{KawNam}, $d$-semi-stability is equivalent
  to the existence of
  a log structure.
\end{enumerate}

Examples of smooth Calabi-Yau three-folds constructed by this
method can be found for example in \cite{KawNam} and
\cite{LeeThesis}.

The proof of the theorem proceeds as follows. For each logarithmic
structure, one can show that the obstruction to a deformation can
be lifted order by order in perturbation theory. Then, the
deformation is `representable' over the ring of formal power
series on the disk.

To show that one has an actual deformation, one proceeds as
follows.%
\footnote{We are grateful to Y.Kawamata for correspondence on this
issue.} In the analytic category, one knows that the Kuranishi
space (i.e.\ the actual analytic deformation space) always exists,
and then the formal power series result above shows that there is an analytic
smoothing of the variety.

In the algebraic category one can make a similar argument, with the additional
assumption that $H^2({\cal O}_{Y_0})=0$.
If one starts with an ample line bundle on $Y_0$, then the assumption
on $H^2$ above implies that the line bundle extends to the analytic
smoothings.  It follows that the family extends to a projective family
by the existence of the Hilbert scheme.  So again one can match the
formal deformation with an actual deformation.

In order to apply the smoothing theorem of Kawamata/Namikawa to
our situation, there are two issues that we should confront. The
first is that Kawamata/Namikawa by itself does not guarantee the
existence of a Weierstrass fibration. This will be addressed in
the next subsection. A second problem is that the
Kawamata/Namikawa theorem assumes the local pieces to be smooth.
On the other hand, we are typically interested in the case that
$W_1$ has interesting singularities, to get non-trivial
non-abelian gauge groups. We may resolve the singularities and
then apply Kawamata/Namikawa, but it is not completely clear that
the exceptional cycles are preserved by the deformation, so that
we can blow them back down.

Let us think about this problem from a physical perspective. The
discriminant locus $\Delta$ of the elliptic fibration of $W_1$
decomposes into two pieces: a multiple of $S_0$ denoted by
$\Delta_0$, and a remainder $\Delta'$ which intersects $S_0$ in
the matter curves. We want $\Delta_0$ to be preserved under the
smoothing. In gauge theory language, this means that the gauge
group should not get Higgsed under smoothing. When $S_0$ is a del
Pezzo, there are no fields in the adjoint representation of the
gauge group, so $\Delta_0$ won't deform by itself. However
$\Delta_0$ might recombine with $\Delta'$. In gauge theory
language, this corresponds to Higgsing the gauge group by the
charged matter located at the intersection of $\Delta'$ with
$S_0$.

So can this happen? Since we are only
considering holomorphic questions here, we should ignore the
$D$-terms. Consider a general superpotential $W$ in the
charged fields $\phi,\tilde\phi$ on $\Delta_0 \cap \Delta'$, and
consider $t$-dependent deformations. The problem comes from the 
$t$-derivative, if we have terms of the form
\be W \ \supset \  P(t) + Q(t){\rm Tr}( \phi \tilde \phi) +
\ldots \ee
which leads to $\phi\tilde\phi \sim P'(t)/Q'(t)$ for small $t$.
However, substituting the VEVs in the derivatives of $W$ with
respect to $\phi,\tilde \phi$, we would get a non-trivial equation
for $t$, whereas we know by Kawamata/Namikawa that there is a
one-parameter family parametrized by $t$. So perhaps $t$-dependent
terms for ${\rm Tr}(\phi\tilde\phi)$ are not generated, and it is
consistent to set the charged fields to zero.

This argument is far from watertight. The question should
be closely related to the choice of log structure. At any rate, we
will see some examples later where we can do the deformation
explicitly, and the gauge group along $S_0$ is preserved.

\newsubsection{Criteria for the existence of an elliptic fibration}

\subseclabel{EllipticCriteria}

Let us assume that the central fiber $Y_0$ carries an elliptic
fibration. In the main case studied in this paper the generic
elliptic fibers are smooth, but this is not necessarily true for
other interesting examples \cite{CDW}. We really want the fibers
$F$ to be connected and satisfy $\chi(F,\cO_F)=0$. We would now
like to see if the smoothing deformation provided by Kawamata and
Namikawa preserves an elliptic fibration structure. This would be
simplest if there were numerical criteria for the existence of an
elliptic fibration on a variety. Such criteria are not known in
complete generality. However there are criteria which cover most
cases of interest, and which are valid in our situation, as was
shown recently in \cite{KollarEllipticReview}.

Let us start with some general remarks. Suppose that a variety $X$
of dimension $d$ has an elliptic fibration $\pi:X \to B$. Consider
an ample line bundle $L_B$ on $B$, and denote $L = \pi^*L_B$.
Since $L_B$ is ample on $B$, for $m$ large enough the space of
sections $H^0(X,L^{\otimes m})$ will give a map from $X$ to
projective space such that the image is isomorphic to $B$. This
encapsulates an important idea in algebraic geometry, that
rational maps often come from divisors in the manner above, and
that we can try to characterize the morphism through the divisor
$L$.  In other words, instead of the data $(X,\pi)$ we will
consider pairs $(X,L)$, from which $\pi$ can be reconstructed. The
projection constructed from $H^0(X,L^{\otimes m})$ with $m$ large
enough is sometimes called the Iitaka fibration associated to $L$.
The fibers of this limiting fibration are connected. Note that as
necessary conditions for $L$ to be of the form $L = \pi^*L_B$, we
have the following numerical criteria: $L\cdot C \geq 0$ for every
curve $C$ on $X$ (i.e. $L$ is nef), $L^d = 0$ and $L^{d-1} \not =
0$.

For the moment let's consider more general fibrations where the fibers
$F$ are still curves, but not necessarily of (arithmetic) genus~1.
Let us consider the holomorphic Euler characteristic
\be \chi(X,L^m) \ = \ \sum_{i} (-1)^i H^i(X,L^m)  \ee
which is a deformation invariant.
Using the
Leray sequence for $\pi:X \to B$ one can show that
\be \chi(X,L^m) \ = \ \chi(F,\cO_F){ m^{d-1}\over (d-1)!}L_B^{d-1}
+  {m^{d-2}\over (d-2)!} L_B^{d-2}\cdot R+ \cO(m^{d-3}), \label{eq:chilm} \ee
where $R$ will be written more explicitly below.  Thus $\chi$
typically grows as $m^{d-1}$.   However, since $\chi(F,{\cal O}_F)
=1-g$, $\chi$ will actually grow to order at most $m^{d-2}$
precisely when the fibers are elliptic. \footnote{It is
interesting to note that by Riemann-Roch, the coefficient of
$m^{d-1}$ in $\chi(X,L^d)$ is given by $(-1/2) L^{d-1}\cdot K_X$,
so the coefficient of $m^{d-1}$ in (\ref{eq:chilm}) vanishes
anyway on a Calabi-Yau. Indeed, by adjunction
$c_1(K_F)=c_1(K_X|_F) + c_1(N_{X/F})$, but for any fibration we
have $N_{X/F}$ is trivial. It follows that $(-1/2)K_X \cdot
\pi^*L_B^{d-1}= (-1/2)\deg((K_X)|_F) L_B^{d-1} = \chi(F,\cO_F)
L_B^{d-1}$, since $\deg((K_X)|_F)=\deg K_F=2g-2 =-2\chi(F,\cO_F)$.
So we conclude that the fibers must be elliptic when $K_X \sim
0$.}

The divisor $R$ on $B$ is given by
\be R\ = \ -\half \chi(F,\cO_F)\, K_B + \sum_i (-1)^i
c_1(R^i\pi_*\cO_X) \ee
Using that $\chi(F,\cO_F)=0$, $R^0\pi_*\cO_X = \cO_B$, and
$c_1(R^1\pi_*\cO_X)= -\Delta_{X/B}$ is (minus) the class of the
discriminant, we can simplify $\chi(X,L^{\otimes m})$ to
\be \chi(X,L^{\otimes m}) \ = \ {m^{d-2}\over (d-2)!}
L_B^{d-2}\cdot \Delta_{X/B} + \cO(m^{d-3}) \ee
Therefore unless the discriminant locus is trivial, we see that
the coefficient of $m^{d-2}$ must actually be non-zero. Using
Riemann-Roch, one further shows that $L_B^{d-2}\cdot \Delta_{X/B}
= L^{d-2}\cdot td_2(X)$ as an expression on $X$, where $td_2(X)$
is defined in terms of a resolution $h:\tilde X \to X$ as
\be td_2(X) \ = \ {1\over 12}h_*\left(c_1(\tilde X)^2 + c_2(\tilde
X)\right) \ee

So modulo some slight subtleties that we have glossed over, the
idea is that we should have an elliptic fibration if we have a nef
divisor $L$ on $X$, with the following properties: $H^0(X,L^m)$
should grow like $m^{d-1}$, so that the Iitaka fibration of $L$
gives a projection $\pi: X \to B$ with connected fibers where $B$
has dimension $d-1$. And $\chi(X,L^m)$ should grow like $m^{d-2}$,
so that the fibers are elliptic. The latter is already a numerical
criterion, and if $L^{d-2}\cdot td_2(X)> 0$ it turns out the
former can be reformulated as one. Namely it is shown in
\cite{KollarEllipticReview} that a pair $(X,L)$ with $L^{d-2}\cdot
td_2(X)> 0$ and $K_X$ nef is an elliptic fiber space if and only
if
\begin{enumerate}
\item $L$ is nef, $L^d=0$ and $L^{d-1} \not = 0$ (the latter modulo torsion);
\item $L^{d-1}\cdot K_X = 0$ and $L - \epsilon K_X$ is nef for $0\leq \epsilon << 1$.
\end{enumerate}
Of course the second condition is automatic if we have $K_X \sim
0$.

Now that we have a good characterization for the existence of an
elliptic fibration, we can ask if it is preserved under a small
deformation. It is worth keeping in mind that this fails in one of
the simplest examples, namely the case of $K3$ surfaces. The
generic deformation of an elliptically fibered $K3$ surface is
certainly not elliptic. As another example, consider the product
of two abelian varieties $A_1 \times A_2$ and let $\pi$ be the
projection on the second factor. The general deformation is a
simple abelian variety which has no projection to a lower
dimensional abelian variety. So there must be an extra condition
if we are to expect an elliptic fibration after a deformation.

The main problems are as follows. First, it is not guaranteed that
$L$ deforms along with $X$. The obstruction lies in
$H^2(X,\cO_X)$, so if we require that $H^2(X,\cO_X)$ vanishes then
this problem is solved.\footnote{If furthermore $H^1(X,\cO_X)=0$ then this
deformation is unique.} This is the extra requirement that fails
for $K3$-surfaces and abelian varieties. Second, even when $L$
deforms, the condition that $L$ is nef is not an open condition in
general. This problem is solved by an induction argument in
\cite{KollarEllipticReview}, which can be made both in the
algebraic and in the analytic category.

We conclude that if $Y_0$ admits an elliptic fibration with
$\Delta_{X/B}\not = 0$ and $H^2(Y_0,\cO_{Y_0})=0$, then the
deformation $Y_t$ provided by Kawamata and Namikawa is also
elliptically fibered. As explained in the introduction, this gives
a new and conceptually very interesting way to construct
$F$-theory compactifications, by assembling the elliptic
Calabi-Yau from more fundamental pieces. More generally, we see
that $F$-theory is rather stable: when $H^2(X,\cO_X)=0$, small
complex structure deformations of $X$ preserve the existence of an
elliptic fibration. This is why the moduli space of an $F$-theory
compactification usually contains an open subset of the full
complex structure moduli space of $X$.

\newsubsection{Example: bubbling off an $SU(5)_{GUT}$ model}

\subseclabel{Example}

To keep things simple and illustrate the ideas, we consider the
following example. We take $Y$ to be an elliptic fibration with
section over $B_3 ={\bf P}^3$, with $I_5$ singularities along a
smooth irreducible divisor $S\subset B_3$ given by  $z=0$. We
explicitly degenerate this as
\be Y_t\ \to\ W_1 \cup_Z W_2 \ee
where $W_1$ is a $dP_9$ fibration over $S$, and $W_2$ is an
elliptic fibration over ${\bf P}^3$ with boundary $Z$, but
different from $Y$. This is one of the simplest examples which
does not have a $K3$-fibration and therefore has no heterotic
dual. See figure 2 for a schematic picture. It is clear that the
same analysis may be applied to other examples, eg. the compact
examples introduced in \cite{Donagi:2009ra}.

\begin{figure}[t]
  \label{DegLimitv1}
\begin{center}
            \scalebox{.8}{
  \includegraphics[width=\textwidth]{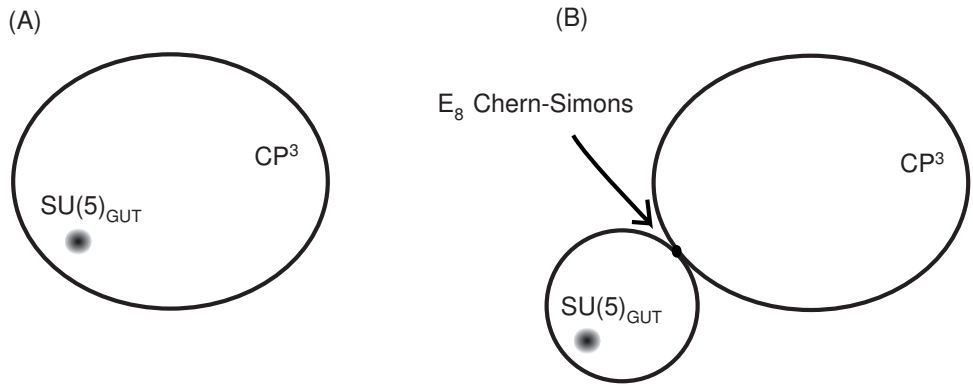}
}
\end{center}
\begin{center}
\parbox{14cm}{ \caption{(A): Global model with $I_5$ singular fibers along a divisor $S$,
corresponding to an $SU(5)_{GUT}$ gauge theory.
 (B): Degenerate version of the global model, in which the singular elliptic fibers describing
 the $SU(5)_{GUT}$ gauge theory have been pushed to a local $dP_9$-fibration.} }
\end{center}
\end{figure}

The elliptic fibration over ${\bf P}^3$ can be written in the Tate form
\be y^2+a_1xy+a_3\ =\ x^3+a_2x^2+a_4x+a_6, \ee
where each of the $a_i$ are homogeneous polynomials of degree $4i$
on ${\bf P}^3$. The $I_5$ condition along the hyperplane $z=0$ says
that $a_i=b_iz^{i-1}$ for each $i$, where $b_i$ is a homogeneous
polynomial of degree $3i+1$ on ${\bf P}^3$. It will be convenient to
write
\be a_i \ = \ b_i z^{i-1} + c_i z^i \ee
There is some ambiguity in such a parametrization, as was the
case in \cite{Katz:2011qp}. In the present
example, we can make it more canonical by requiring $b_i$ to be
independent of $z$.

Applying our results from Section~\ref{LocalDegeneration},
we can achieve the normal cone
degeneration by simply inserting a factor of $t$ in front of each
$b_i$ (since $i-n_i=1$ for all $i$ in the case of $I_5$).  Doing
so and changing the signs of $a_1$ and $a_3$ for convenience, we
have the following elliptic fibration which extends to the normal
cone degeneration:
\be y^2\ =\ x^3 + a_1(t) xy  + a_2(t) z x^2 + a_3(t) z^2 y +
a_4(t) z^3 x + a_6(t)\, z^5. \label{tatesu5} \ee
where
\be a_i(t) \ = \ t b_i z^{i-1} + c_i z^i \ee
For $t\ne0$, there is
clearly an $I_5$ along $z=0$, and $a_i(t)$ vanishes to order $i$
along $z=t=0$ as required. Applying our previous results, at
$t=0$ we get $W_1$ to be the local model with $I_5$ singularities.

Let's look at the limit in the neighborhood of the intersection of
$W_1$ and $W_2$.   In the notation of Section~\ref{localglobal} we
have the total space $\tilde{\cal B}$ of the normal cone
degeneration, and for $t=0$ we have two components.  In the
current situation, one component (the base of $W_2$) is
isomorphic to ${\bf P}^3$, and the other component (the
base of $W_1$) is isomorphic to the projective bundle ${\bf
P}(\cO\oplus\cO(1))$ over ${\bf P}^2$. Local coordinates near the
intersection of these two components include $z$ and $(t/z)$, so
that $t=0$ has two components because $t=z(t/z)$. The component
$z=0$ is part of the exceptional divisor $E$ and $t/z=0$ gives
${\bf P}^3$.

To pass to the elliptic fibration over the normal cone
degeneration, we replace the canonical bundle of ${\bf P}^3$ by
the relative dualizing sheaf $\omega_\pi$. The effect of this is
to require the removal of $i$ copies of the exceptional divisor
from each $a_i(t)$. In our local coordinates, this means that we must
replace each $a_i(t)$ by $\hat{a_i}=a_i(t)/z^i$, so that
\be \hat{a_i}=\frac{t}{z}b_i+c_i. \ee
Thus, (\ref{tatesu5}) becomes after the normal cone degeneration
\ba\label{CentralFibEqn} y^2& =& x^3 + \frac{t}{z} \left(b_1 xy  + b_2 x^2 +
b_3  y + b_4 x+b_6\right)\eol & & \qquad + \left(c_1 xy  + c_2  x^2 + c_3 y+
c_4 x+c_6\right) \ea
We restrict to the ${\bf P}^3$ component by putting $t/z=0$. This
gives us $W_2$:
\be\label{W2Eqn} y^2\ =\ x^3 + c_1 xy + c_2  x^2 + c_3 y + c_4
x+c_6. \ee
The normal crossing divisor $Z=W_1 \cap W_2$ is given by further
setting $z=0$ in this equation. Note that the $c_i$ are sections
of $(K_{B_3} \otimes \cO(S))^{-i}$, so $c_i|_{z=0}$ are sections
of $K_S^{-i}$ by adjunction. Thus $Z$ is an elliptic Calabi-Yau
three-fold. Equation (\ref{W2Eqn}) is simply a generic elliptic
fibration $W_2$ over ${\bf P}^3$ with the property that $(W_2,Z)$ is a
log Calabi-Yau fourfold.

By setting $z=0$ in (\ref{CentralFibEqn}) and using $t/z$ as a
local coordinate, we are describing a patch of $W_1$ near the
infinity section, where it is glued onto $W_2$. The $I_5$
singularities are located at the zero section $S_0$, which is
disjoint from the infinity section $S_\infty$. To see them
explicitly, we rescale by $z/t$ and use $\tilde{z}=z/t$ as a local
coordinate as we did in (\ref{eq:ellipticw1}) where the more general
$s$ was used instead of $z$. We get
\ba y^2& =& x^3 + b_1 xy + b_2 \tilde{z} x^2 + b_3 \tilde{z}^2 y + b_4 \tilde{z}^3
x+b_6 \tilde{z}^5\eol & & \qquad + (c_1 \tilde{z} xy  + c_2 \tilde{z}^2 x^2 + c_3 \tilde{z}^3
y+ c_4 \tilde{z}^4 x+c_6 \tilde{z}^6). \ea
Here, all the $b_i$ and $c_i$ are evaluated at $z=0$. We recognize
the typical form of a $dP_9$-fibration with $I_5$ singularities at
$\tilde{z}=0$.

\newsubsection{The general case}

We return to the general case.
Suppose conversely we are given two local models, i.e.\ two elliptically
fibered log Calabi-Yau fourfolds $(W_i,Z)$ for $i=1,2$ which we glue along
their common elliptically fibered boundary $Z$ to form $Y_0=W_1\cup_Z W_2$.
We want to know when this model corresponds to an F-theory limit, i.e.\
when $W_1\cup_Z W_2$ has a smoothing to an elliptically fibered Calabi-Yau
fourfold $Y_t$.

As discussed earlier, we can find a smoothing when the
Kawamata-Namikawa conditions hold.  Let $B_i$ be the base of $W_i$ and
let $S$ be the base of $Z$, so that $B_0=B_1\cup B_2$ is the base of $Y_0$.
We must assume that $H^1(\cO_{W_i})=0$ (for which it suffices to take
$B_i$ with $H^1(\cO_{B_i})=0$ by the Leray spectral sequence), and the assumption
that $Y_0$ is K\"ahler is easy to achieve in practice.  The only other
obstruction to the smoothability of $Y$ is the $d$-semistability condition
$N_{Z,W_1}\otimes N_{Z,W_2}\simeq\cO_Z$, which we now assume.  With these
assumptions, $Y_0$ smooths.  If we further
assume that the discriminant $\Delta_{Y_0/B_0}$ is non-empty and
$H^2(\cO_{Y_0})=0$, then the smoothing is elliptically fibered and we are done.
We will return to these general considerations later, but we digress slightly
to specialize a bit.

As we have already discussed, the most interesting situation for
$F$-theory occurs when one of the local models
(say $W_1$) has a $dP_9$ fibration
as well as a prescribed singularity along a divisor $S_0\subset B_1$
disjoint from $S$.  We now claim that these assumptions imply that
$B_0$ coincides with the limit $B_0$ of the degeneration to the normal
cone of $S\subset B_2$, and we are back in the situation illustrated
by examples in Section~\ref{LocalDegeneration} and
Section~\ref{Example}.  Even still, we have gained something, as the
more abstract Kawamata-Namakawa method makes it clear that our earlier
assumptions of Tate form were not essential, so that the smoothing can
be done if the prescribed singularities are given in a more general way.
Furthermore, we can relax
the hypotheses in the previous paragraph (for example by dropping the
requirement of a heterotic dual) and achieve additional F-theory limits
with more general bases $B_0$.

We now prove our claim.  The assumption is that
$B_1$ is fibered over $S$ by ${\bf P}^1$s,  with disjoint sections $S$
and $S_0$.  This implies that $B_1$ is the projectivization of a decomposable
bundle $L_1\oplus L_2$, hence is isomorphic to ${\bf P}(\cO_S\oplus L)$ for a line
bundle $L$ on $S$, with either $L=L_2\otimes L_1^{-1}$ or
$L=L_1\otimes L_2^{-1}$.
Since $B_1-S_0$ is the total space of $N_{S/B_1}$, we see that $B_1$
must therefore be
${\bf P}(\cO_S\oplus N_{S/B_1})$.  Next, since $N_{Z/Y_i}$ is the pullback
of $N_{S/B_i}$ by the elliptic fibration, the $d$-semistabity condition is
that  the pullback of $N_{S/B_1}\otimes N_{S/B_2}$ is trivial.  However,
since $\mathrm{Pic}(S)\to\mathrm{Pic}(Z)$ is injective, we conclude that
$N_{S/B_1}\simeq (N_{S/B_2})^{-1}$.  Then
$B_2={\bf P}(\cO_S\oplus N_{S/B_1})\simeq {\bf P}(\cO_S\oplus N_{S/B_2}^{-1})
\simeq {\bf P}(N_{S/B_2}\oplus \cO_S)$.  Note that $N_{S/B_2}$ is now
identified with a
neighborhood of $S_0$ and we have achieved the limit
of the degeneration to the normal cone.

We now return to the general case and be more explicit, using
Weierstrass form.  We start
with projective bases
$B_1$ and $B_2$ with $H^1(B_i,\cO_{B_i})=H^2(B_i,\cO_{B_i})=0$
sharing a common divisor $S$ with inverse
normal bundles in the respective components.  Let $K_i=K_{B_i}+S$ be the
restriction of the dualizing sheaf of $B_0$ to $B_i$, and assume that
$-4K_i$ and $-6K_i$ have enough sections $f_i\in H^0(B_i,-4K_i)$
and $g_i\in H^0(B_i,-6K_i)$ with $(f_1)|_S=(f_2)|_S$ and $(g_1)|_S=(g_2)|_S$
so that we have a Weierstrass fibration
%
\be y^2 \ = \ x^3 + fx + g, \ee
where $f$ and $g$ are the sections on $B_0$ obtained from the gluing.


Let's see what else we need for all the necessary conditions.
The first condition was that the pairs $(W_1,Z)$ and $(W_2,Z)$
should be log Calabi-Yau spaces, but that is automatic since
$f_i$ and $g_i$ were constructed from the log canonical bundles.

The remaining conditions are also easy to check or achieve.  We have already
checked that the $d$-semistability condition holds since it pulls back
from the base.  In order for $B_0$ to be projective, we need to find
ample line bundles $L_i$ on $B_i$ which agree  on $S$.  This is easy to
achieve if the restrictions of $Pic(B_i)$ to $Pic(S)$ are big enough; or
we can just assume it.  And
there is nothing to check in the case of the normal cone degeneration since
the degeneration can be achieved within projective geometry by blowups as
we have seen.  Thus, with the above mild assumptions, $Y_0$ deforms.

Finally, we want to what conditions are needed so that the deformation is elliptic.  Since we are using
Weierstrass form,
we only need generic enough sections $f,g$
for the discriminant $\Delta=4f^3+27g^2$ to be a nontrivial divisor, so we
assume that.
Furthermore, $H^2(\cO_{Y_0})=0$
by a Leray spectral sequence argument.

Then we may apply the results of \cite{KollarEllipticReview}
discussed in Section \ref{EllipticCriteria}, and $Y_0$ deforms to
an elliptically fibered Calabi-Yau fourfold, with a built-in
degeneration limit.

\newpage

\newsection{The form of the Lagrangian in a degeneration limit}

\seclabel{VHS}

As we explained, we are not really interested in complex spaces
themselves, but in holomorphic field theories on such spaces. Such
holomorpic field theories lead to interesting moduli spaces, and
differential equations on bundles over these moduli spaces.
Degeneration limits give rise to boundaries on such moduli spaces,
and we now want to study how correlators (solutions of
differential equations) behave near a boundary.

Let us think back for example about the case of conformal blocks
of CFTs on a Riemann surface. The conformal blocks of a CFT are
mathematically described by (twisted) $D$-modules, which are
roughly speaking holomorphic vector bundles over the moduli space
with a (projectively) flat connection. The connection usually
develops a logarithmic singularity when the Riemann surface
degenerates ($\nabla \sim d + L_0 dt/t$), and the monodromies give
an action of the mapping class group on the correlators. This
action plays an important role in two-dimensional CFT.

One expects similar phenomena for holomorphic field theories in
higher dimensions, and $F$-theory provides another case study. In
the large volume limit of $F$-theory the moduli space is a
fibration, with base given by the complex structure moduli space
of the Calabi-Yau and fiber given by the intermediate Jacobian
${\cal J}^2 = H^3_{\bf C}/(F^2H^3_{\bf C} + H^3_{\bf Z})$. In what
follows we focus on the complex structure moduli and ignore the
Jacobian. We get a $D$-module structure from the Gauss-Manin
connection on the Hodge bundle over the complex structure moduli
space of the Calabi-Yau.

In this section we want to understand the behaviour in a
semi-stable degeneration limit. Fortunately the general aspects of
this problem are well-understood. In appropriate coordinates, near
the degeneration limit the Gauss-Manin connection takes the form
\be \nabla \ = \ d + N {dt\over t} \ee
where $N$ is a nilpotent matrix. Once we know the matrix $N$, we
can use this to make qualitative statements about the behaviour of
the effective lagrangian without having to calculate a single
period integral. As we will see, the matrix $N$ can be deduced
from the geometry of the central fiber. Thus this fits very well
with the cutting-and-gluing approach.

\newsubsection{Asymptotics from monodromy}

\subseclabel{SchmidOrbit}

On a smooth fourfold, the superpotential for the supergravity
fields is the flux superpotential, which appears to take the form
of a period integral:
\be\label{FluxW} W \ =\ {1\over 2\pi}\int_Y \Omega^{4,0} \wedge
{\sf G}. \ee
with ${\sf G}/2\pi-1/2 p_1(TY)$ having integral periods.
Thus we want to understand
the behaviour of periods in a semi-stable degeneration limit,
which is a classical problem in Hodge theory.

We have to make some caveats. For phenomenological applications we
are not quite interested in smooth Calabi-Yau spaces, but in
situations where certain exceptional cycles are contracted, in
order to get non-abelian gauge symmetries. The above
superpotential does not capture interactions of quantized $M2$
solitons wrapped on exceptional cycles, since such degrees of
freedom are much heavier and are not obtained by KK reduction from
supergravity. Such degrees of freedom however become important in
the limit that the exceptional cycles shrink to zero, and there
exist branches of non-abelian solutions which are not captured by
(\ref{FluxW}). Nevertheless we expect the $t$-dependent
corrections to be qualitatively similar.

Further, the superpotential is often thought of as being defined
on the infinite dimensional space of field configurations, i.e. it
also depends on all the massive KK modes. Here we are only
evaluating it on harmonic forms. Still let us see what we can
learn from only a few general considerations.

The qualitative form of the periods in a degeneration limit is
rather constrained, and can be understood using Hodge theory. We
have given a brief review of the relevant material in appendix
\ref{MHS}. Here we will be interested in general properties of the
case of a semi-stable degeneration $\pi_{{\cal Y}}:{\cal Y} \to
\Delta$, where $Y_t$ is a smooth Calabi-Yau fourfold for $t\not =
0$, and $Y_0 = W_1 \cup_Z W_2$, with $(W_1,Z)$ and $(W_2,Z)$ log
Calabi-Yau spaces and $Z$ a Calabi-Yau threefold.

The lattice $H_4(Y_t,{\bf Z})$ forms a local system over the
$t$-plane away from $t=0$. We can use parallel transport to get an
isomorphism with $H_4(Y_{t'},{\bf Z})$ for $t' \not = t$. Now let
us circle around the origin of the $t$-plane. Then the cycles in
$H_4(Y_t,{\bf Z})$ get rearranged, and this is expressed as an
automorphism $M$ of $H_4(Y_t,{\bf Z})$ called the monodromy
transformation. It is known in general that up to a base change
(i.e. reparametrizing $t$ as $t \sim \tilde t^k$), the monodromy
transformation is unipotent. In our situation with a semistable
degeneration, the monodromy was already unipotent and so a base
change is not necessary. We also introduce the logarithm $N =
\log(M)$, which is nilpotent when $M$ is unipotent.

Let us collect the periods of $\Omega^{4,0}$ with respect to some
basis of $H_4(Y_t,{\bf Z})$ in a vector $\vec{\Pi}(t)$. Now the
Schmid nilpotent orbit theorem says that as $t \to 0$ the periods
are approximated in the following way:
\be\label{PeriodLimit} \vec{\Pi}(t)\ \sim\  e^{{1\over 2\pi i}N
\log(t)}\vec{\Pi}_0  \ee
where $N$ is the logarithm of the monodromy matrix, and
$\vec{\Pi}_0$ is independent of $t$. The expression on the right,
i.e. the nilpotent orbit, may be thought of as a perturbative
approximation to the periods. Thus we want to know the matrix $N$
and we want to know $\vec{\Pi}_0$. In other words, we want to know
the limiting mixed Hodge structure associated to the degeneration
$t\to 0$. In the following, we delve into a slightly lengthy
analysis in order to derive this. Readers who might not want to
follow this derivation may skip to equation (\ref{WNilp}), where
we write the form of the superpotential that follows from our
analysis.

\newsubsection{The limiting mixed Hodge structure}

\subseclabel{LimMHS}

As reviewed in appendix \ref{MHS}, there are at least two ways for
getting the limiting mixed Hodge structure. The approach we will
use below is the Clemens-Schmid exact sequence, which allows us to
compare the limiting mixed Hodge structure for $t \to 0$ with the
mixed Hodge structure of the central fiber $Y_0$. An alternative
approach, also briefly mentioned in appendix \ref{MHS}, is to
study the mixed Hodge structure of the logarithmic cohomology
groups $H^k_{\rm log}(Y_0)$ of the central fiber.

The part of the Clemens-Schmid sequence that we will use is
\be\label{cs} \cdots\to H_{6}(Y_0)\ \stackrel{\alpha}{\to}\
H^4(Y_0)\ \stackrel{i^*}{\to}\ H^4_{\rm lim}(Y_t)\
\stackrel{N}{\to}\ H^4_{\rm lim}(Y_t)\ \stackrel{\beta}{\to}\
H_{4}(Y_0)\ \to\ \cdots \ee
Since $Y_0$ is a deformation retract of ${\cal Y}$, we see that
the homologies and comologies of $Y_0$ are identified with those
of ${\cal Y}$.  Then $i^*$ is identified with the usual pullback
$i^*:H^k({\cal Y})\to H^k(Y_t)$.  The maps $\alpha$ and $\beta$
are induced by Poincar\'e duality on ${\cal Y}$ and $Y_t$
respectively. All of the terms in the Clemens-Schmid sequence
carry natural mixed Hodge structures. The maps $\alpha,\ i^*,\ N$,
and $\beta$ are morphisms of mixed Hodge structures, shifting the
degrees by $(5,5),\ (0,0),\ (-1,-1)$, and $(-4,-4)$ respectively.

Now we need the weight filtration for $H^4(Y_0)$. There is a
general prescription which can be found in appendix \ref{MHS}, but
in the present case we can easily read it off from the
Mayer-Vietoris sequence for $Y_0$, which is given by
\be\label{MVCohomologyLES} \ldots \to  H^3(W_1) \oplus H^3(W_2)\
\mathop{\to}^{{\sf d}^3} \ H^{3}(Z)  \to  H^4(Y_0)  \to  H^4(W_1)
\oplus H^4(W_2) \ \mathop{\to}^{{\sf d}^4} \ H^4(Z)  \to  \ldots
\ee
This gives a short exact sequence
\be 0\ \to\ {\rm coker}({\sf d}^3)  \ \to\ H^4(Y_0)\ \to\  {\rm
ker}({\sf d}^4)\ \to\ 0\ee
and so the weight filtration on $H^4(Y_0)$ is simply given by
\be 0\ \subseteq\ \W_3\ \subseteq\ \W_4=H^4(Y_0) \ee
with $\Gr_4H^4(Y_0)=\ker({\sf d}^{4})$ and $\Gr_3H^4(Y_0)={\rm
coker}({\sf d}^{3})$.

The Clemens-Schmid sequence (\ref{cs}) induces exact sequences on
the graded pieces.  The Mayer-Vietoris sequence on homology is
given by
\be\label{MVHomologyLES} \ldots \to   \ H_k(W_1) \oplus H_k(W_2)\
  \to\  H_k(Y_0)\  \to\  H_{k-1}(Z)  \ \to \ldots \ee
This shows that the non-zero graded pieces of $H_6(Y_0)$ are of
weights $-6$ and $-5$, while those of $H_4(Y_0)$ are of weights
$-4$ and $-3$.

Since $\Gr_{-7}H_6(Y_0)=0$ and $\Gr_1H^4(Y_t)=0$, the relevant
graded piece of (\ref{cs}) gives an isomorphism
\be \Gr_3H^4(Y_0)\ \cong\ \Gr_3H^4_{\rm lim}(Y_t)\ee
so that
\be \Gr_3H^4_{\rm lim}(Y_t)\ \simeq\ {\rm coker}({\sf d}^{3}).\ee
We also have the isomorphism
\be \Gr_5H^4_{\rm lim}(Y_t)\ \stackrel{N}{\simeq}\ \Gr_3H^4_{\rm
lim}(Y_t)\ee
by a general property of the monodromy weight filtration, or
alternatively from another graded piece of (\ref{cs}).

Let $p$ be the smallest integer such that $\Gr_pH^4(Y_t)\ne0$. We
claim $p=3$.  If $p<3$, we have from (\ref{cs}) the short exact
sequence
\be \Gr_pH^4(Y_0)\ \to\ \Gr_pH^4(Y_t)\ \stackrel{N}{\to}\
\Gr_{p-2}H^4(Y_t)\ee
Since the left and right terms vanish by our assumptions, it
follows that $\Gr_pH^4(Y_t)=0$, a contradiction. Because
$N^k:\Gr_{4+k}{\to}\Gr_{4-k}$ is an isomorphism,
we see that the only non-zero graded pieces of
$H^4(Y_t)$ are $\Gr_3$, $\Gr_4$ and $\Gr_5$, and the matrix $N$
satisfies $N^2 = 0$. Since the weight filtration is non-trivial,
we also have $N\not = 0$. More generally this argument shows that
if the central fiber only has strata up to codimension $k$, so
that if $\W_{n-m}H^n(Y_0) = 0$ for $m> k$, then we will have
$N^{k+1}=0$.

To summarize, we found that $N^2=0$ but $N \not = 0$. The
monodromy weight filtration on $H^4(Y_t)$ therefore has the form
\be\label{MHSFlag} 0 \ \subseteq\ \W_3 \ \subseteq \ \W_4 \
\subseteq \ \W_5 = H^4_{\rm lim}(Y_t) \ee
where $\W_3H^4(Y_t) = {\rm im}(N)$ and $\W_4H^4(Y_t) = {\rm
ker}(N)$. We also found the following isomorphisms of Hodge
structures:
\be \Gr_5H^4_{\rm lim}(Y_t)\ \stackrel{N}{\simeq}\ \Gr_3H^4_{\rm
lim}(Y_t)\ \simeq\ \Gr_3H^4(Y_0)\ \simeq\ {\rm coker}({\sf
d}^{3}).\ee
One can say more than this. From (\ref{cs}) we also see that
\be  \Gr_4H^4_{\rm lim}(Y_t)\ \simeq\ \Gr_4H^4(Y_0)/{\rm ker}(i^*)
\ \simeq\ {\rm ker}({\sf d}^{4})/{\rm ker}(i^*) \ee
as well as
\be \W_4H^4_{\rm lim}(Y_t)\ \simeq\  \W_4H^4(Y_0)/{\rm ker}(i^*) =
H^4(Y_0)/{\rm ker}(i^*) \ee
Now the map $\alpha$ consists of Poincar\'e duality
$H_6(Y_0) \cong H_6({\cal Y}) \to H^4({\cal Y}, \del{\cal Y})$, which
is an isomorphism, followed by the natural map
$H^4({\cal Y}, \del{\cal Y}) \to H^4({\cal Y})\cong H^4(Y_0)$.
Furthermore, by the Thom isomorphism we have $H^4({\cal Y}, \del{\cal Y})
\cong H^2(Y_0)$. Then the geometric picture for
${\rm ker}(i^*)={\rm im}(\alpha)$ is as follows. When we
glue $W_1$ and $W_2$, we have to make certain identifications on
the cohomology, as we can see from (\ref{MVCohomologyLES}).
Similarly we have to make certain identifications in homology.
From (\ref{MVHomologyLES}) we see that when a four-cycle in $Z$
embeds non-trivially in $W_1$ and $W_2$, then it has to be
identified in $Y_0$. Now the Poincar\'e duals of these
classes in $W_1$ and $W_2$ are not identified when we glue; they
descend to two distinct cohomology classes in $Y_0$. But we expect
that they should become equivalent when we deform to $Y_t$, since
$Y_t$ is smooth and therefore has Poincar\'e duality. As a result
a linear combination of these classes is in the kernel of $i^*$.
We can see this a bit more easily using the logarithmic approach,
using equation (\ref{LogResLES}).

As we will see in a bit more detail in the next section, the
remaining information in the flag structure (\ref{MHSFlag}) can
also be extracted. We can now use this knowledge of the monodromy,
together with the nilpotent orbit theorem, to compute the
asymptotics of periods in the degeneration limit.

\newsubsection{Asymptotic form of the superpotential}

\subseclabel{WAsymptotics}

The nilpotent orbit theorem says that the degeneration of Hodge
structures behaves asymptotically like $\mathrm{exp}({1\over 2\pi
i}\log(t)N)F_{\rm lim}^\bullet$, where $F_{\rm lim}^\bullet$ is
the Hodge filtration of the limiting mixed Hodge structure on
$H^4(Y_t)$. For the holomorphic four-form, we are interested in
the $F^4$ component. Let $\Omega_0\in F^4$ be a nonzero element of
the one-dimensional vector space $F^4$.  Then $N\Omega_0\in
F^3\cap \W_3$ is nonzero. This gives, since $N^2=0$
\be \Omega(t)\ \sim\ \Omega_0(t)\ \equiv\ \Omega_0+\frac1{2\pi
i}\log(t)N\Omega_0\ee
In order to find the superpotential, we need to write the period
map. Let us take a basis $\vev{e^i,f^j,g^k}$ for $\W_5 =
H^4(Y_t,{\bf Z})$ which is adapted to the monodromy weight
filtration. That is, the $e^i$ projects to a basis of $\W_5/\W_4$,
the $f^j$ project to a basis $\W_4/\W_3$, and the $g^k$ span
$\W_3$. The matrix $N$ acts as $N e^i = g^i$, $Nf^j = N g^k = 0$.
We also consider the dual basis $\vev{e_i,f_j,g_k}$ for
$H_4(Y_t,{\bf Z})$. Then we can decompose $\Omega_0$ as
\be\label{OmZeroPeriod} \Omega_0 \ = \ e^i \int_{e_i}\Omega_0 +
f^j\, \int_{f_j}\Omega_0 + g^k\,  \int_{g_k} \Omega_0 \ee
and we can decompose the $F^4$ part of the nilpotent orbit as
\be\label{OmtPeriod} \Omega_0(t) \ = \ e^i \int_{e_i}\Omega_0 +
f^j\, \int_{f_j}\Omega_0 + g^k\, \left( {1\over 2\pi
i}\log(t)\int_{e_k}\Omega_0+ \int_{g_k} \Omega_0\right)\ee
Using the basis $\vev{e^i,f^j,g^k}$ to fix an isomorphism
$\Pi:H^4(Y_t,{\bf Z}) \cong {\bf Z}^{\dim H^4}$, we can think of
this as the period map. As our decomposition suggests, we can
understand the properties of $\Omega_0$ in $F^4\cap \W_5 =
F^4H_{\rm lim}^4(Y_t)$ by successively building up $\W_5$ from the
pure Hodge structures on ${\Gr}_5$, ${\Gr}_4$ and ${\Gr}_3$ using
the weight filtration.

First we consider the image of $\Omega_0$ in $F^4 {\Gr}_5$, by 
which we mean the projection $(F^4+\W_4)/\W_4$.
Here we can use the fact that $F^4{\Gr}_5\simeq F^3{\rm
coker}({\sf d}^3)$ is an isomorphism. Define $\vev{d^i}$ to be a
basis for ${\rm coker}({\sf d}^3)$ which gets mapped to
$\vev{e^i}$ and $\vev{g^i}$ under the isomorphisms $\Gr_5 \cong
\Gr_3 \cong {\rm coker}({\sf d}^3)$, and let $\vev{d_i}$ be its
dual. Under the isomorphism we have
\be \int_{e_i} \Omega_0 \ = \ \int_{d_i} \Omega^{3,0} \ee
This is the expression we will use for the first term in
(\ref{OmZeroPeriod}).

In the next step, we try to lift this to $F^4(\W_5/\W_3)= (F^4+\W_3)/\W_3$.
Note that from the short exact sequence $0 \to \W_4 \to \W_5 \to
{\Gr}_5 \to 0$, it follows that $\W_5/\W_3$ sits in the following
short exact sequence of mixed Hodge structures:
\be 0 \ \to\ {\Gr}_4 \ \to \ \W_5/\W_3 \ \to \ {\Gr}_5 \ \to \ 0
\ee
and we are interested in the $F^4$ part. We have already discussed
$F^4{\Gr}_5$, and in fact we have $F^4{\Gr}_4 = (F^4\cap \W_4 + \W_3)/\W_3=0$. 
To see this, ${\Gr}_4$ has a weight four Hodge structure so could a
priori have subspace of Hodge type $(4,0)$. However,
$\Gr_4H^4(Y_t)$ is a quotient of $\ker({\sf d}^{4})\subset
H^4(W_1)\oplus H^4(W_2)$. Since $W_1$ and $W_2$ are merely log
Calabi-Yau fourfolds, they have no holomorphic four-forms, so
indeed $F^4 {\Gr}_4 = 0$. For future use, also note that
$F^4\cap \W_3=0$, as $\Gr_3$ has a Hodge structure of weight three
which trivially has no $F^4$ part. It then also follows that
$F^4\cap \W_4=0$.

Now even though $F^4 {\Gr}_4 = 0$, it is not true that
$\int_{f_j}\Omega_0$ vanishes. The reason is that $\W_5/\W_3$ is
not simply a sum of graded pieces, but rather a non-trivial
extension of mixed Hodge structures. So to recover $F^4
(\W_5/\W_3)$, we also need to study this extension class, denoted
by $\Ext^1_{\MHS}({\Gr}_5,{\Gr}_4)$.

Let us see this more explicitly for one of the simplest possible
extensions, namely the extension of the constant Hodge structure
${\bf Z}(0)$ by ${\bf Z}(1)$. Explicitly, we have $H_{\bf Z} =
\vev{h^0,h^1}$. The weight filtration is given by $\W_{-2}H =
\vev{h^0} \subseteq W_0 = H$, and the Hodge filtration is given by
$F^0H_{\bf C} = \vev{h^1-\log(q) h^0}\subseteq H_{\bf C}$ for some
$q \in \Ext^1_{\MHS}({\bf Z}(0),{\bf Z}(1)) ={\bf C}^*$. Note that
$F^0 \cap \W_{-2} = 0$, much like $F^4\Gr_4 = 0$ above, since
$F^0$ and $\W_{-2}$ are two distinct planes in ${\bf C}^2$, which
intersect only at the origin. Nevertheless, if in analogy with
above we denote $h^1-\log(q)
 h^0$ by $\Omega$, then we could write the period map as
\be \Omega \ = \ h^1 \int_{h_1} \Omega + h^0 \int_{h_0} \Omega \ee
and we have $\int_{h_0}\Omega = -\log(q) \not = 0$.

The problem of characterizing extensions of mixed Hodge structures
has been studied by Carlson \cite{CarlsonMHSExt}. Let us consider
a general extension
\be 0\ \to\  A\  \to\  C\ \to\ B\ \to\ 0 \ee
If the mixed Hodge structure is separated, in the sense that the
highest weight of $A$ is less than the lowest weight of $B$, then
the generators of $\Ext^1_{\MHS}(B,A)$ are parametrized by the
generalized complex torus
\be  {\cal J}^0\Hom(B,A)\ \equiv \ \Hom_{\bf C}(B,A)/(F^0\Hom(B,A)
+ \Hom_{\bf Z}(B,A)) \ee
Here $F^0\Hom(B,A) = \{\phi \in \Hom_{\bf C}(B,A) | \phi(F^p B)
\subseteq F^pA \}$. Furthermore, in geometric situations we can
usefully represent these maps as follows. Let $\sigma_{\bf C}: B
\to C$ be a section, i.e. a map such that $\sigma(F^pB) \subseteq
F^pC$ but which does not preserve the integral structure. Let
$r_{\bf Z}:C_{\bf Z}\to A_{\bf Z}$ be an integral retraction,
which is defined as follows: take an integral basis $e^i$ for $A$,
consider the dual basis $e_i$ for $A^\vee$, and lift it to
$C^\vee$ using the surjection $C^\vee \to A^\vee$. Then we have
$r_{\bf Z}(c) = \sum_i\vev{c,e_i} e^i$. This preserves the
integral structure but not the Hodge structure. Using these maps,
the extension classes may be represented as $\psi = r_{\bf Z}
\circ \sigma_{\bf C}$. In the above example, this would give
$\psi(\Omega) = h^0 \int_{h_0} \Omega= -\log(q) h^0$.

We want to apply this result to our short exact sequence:
\be\label{W3W5Ext} 0 \ \to \ {\rm Gr}_4 \ \to \ \W_5/\W_3 \ \to \
{\rm Gr}_5 \ \to \ 0 \ee
which is certainly separated. Let us first discuss an important
special case, which arises in heterotic/$F$-theory duality (see
\cite{Curio:1998bva} and appendix C of \cite{Donagi:2008ca}) and
in IIB/$F$-theory duality \cite{CDW}. In these cases we have
certain special divisors (``cylinders'') $R_{1,2} \subset
W_{1,2}$, intersecting the boundary in $R_{1,2} \cap Z = C_{1,2}$,
which can be used to construct an isomorphism of Hodge structures
appearing on the two sides. In the $E_8\times E_8$ case, we get an
isomorphism
\be H^2(C_1)_\Lambda \ \to\  H^4(W_1)_\Lambda\ee
and similarly for $W_2$. Here $H^4(W_1)_\Lambda \subset H^4(W_1)$
is the sublattice corresponding to the allowed ${\sf G}$-fluxes in
$F$-theory, and $H^2(C_1)_\Lambda$ is a sublattice which is
defined as follows:  we have an action of the $E_8$ Weyl group on
$C_1$, and the cohomology of $C_1$ decomposes into isotypic
pieces. The sublattice $H^2(C_1)_\Lambda$ corresponds to the
eight-dimensional representation, obtained from the action of the
Weyl group on the $E_8$ root lattice. To descend to $\Gr_4$, we
further have to restrict this to ${\rm ker}({\sf d}^4)$. Such
classes end up in $H^2(C_1)_{\rm van}$, which is defined to be the
orthogonal complement (with respect to the intersection pairing)
of the image of the restriction mapping $H^2(Z) \to H^2(C_1)$. In
other words, it consists of those classes whose Poincar\'e dual in
$H_2(C_1)$ becomes homologically trivial when embedded in $Z$.
There is a very similar story for the $SO(32)$ heterotic and IIB
limits \cite{CDW}.

As before, we also have the isomorphism ${\Gr}_5 \cong {\rm
coker}({\sf d}^3)$. In heterotic/$F$-theory duality and
IIb/$F$-theory duality, this is actually simply $H^3(Z)$. So
apparently, our short exact sequence (\ref{W3W5Ext}) is describing
an extension of $H^3(Z)$ by a subgroup of $H^2_{\rm
van}(C_{1,2})$.

Now there is indeed such an extension, and its $F^4$ part (or
$F^3$ part after the isomorphisms, which shift the degree down)
was described in some detail in section 9 of \cite{ClemensDef}.
Namely if $H^3(C) = 0$, which holds for example if $C$ is very
ample, we have the following exact sequence for the relative
cohomology group $H^3(Z,C)$:
\be 0 \ \to\ H^2_{\rm van}(C) \ \to \ H^3(Z,C)\ \to \ H^3(Z) \to \
0 \ee
Further restricting to $H^2(C)_\Lambda$, as we will implicitly
assume in the following, the extension yields a subgroup of
$H^3(Z,C)$. Then we can apply Carlson's prescription to get an
explicit expression for the extension class. We take an integral
basis $c^j$ for $H^2_{\rm van}(C)$ mapping to $f^j$ in $\Gr_4$,
and denote the dual basis of $H_{2,{\rm van}}(C)$ by $c_i$. A lift
to $H_3(Z,C)$ is given by choosing a set of three-chains
$\Gamma_j$ such that $\del \Gamma_j = c_j$. Note that such a lift
is ambiguous up to classes in $H_3(Z,{\bf Z})$. We further take
the holomorphic three-form $\Omega^{3,0}$ in $H^3(Z)$ and lift it
to $H^3(Z,C)$, which can be done since it vanishes in $H^2(C)$.
Then the homomorphism representing the $F^3$ part of the extension
class is given by
\be \psi \ = \ \sum_j c^j \, \psi_j \ = \ \sum_j c^j
\int_{\Gamma_j} \Omega^{3,0} \ee
We recognize this as one of the ways to write the superpotential
for $7$-branes wrapped on $C\subseteq Z$, with worldvolume gauge
flux given by $c^j$. In general $C$ is not smooth, but we will ignore
the singularities here. It is shown in \cite{ClemensDef} that when
$C$ is smooth, $\psi$
(or more precisely, any integral linear combination of the
individual terms) has all the expected properties: it varies
holomorphically in the complex structure moduli, and its critical
locus (with respect to the complex structure moduli that keep $Z$
fixed) is the Noether-Lefschetz locus, where the worldvolume gauge
flux is of Hodge type $(1,1)$. Up to a shift by a class inherited
from $H^2(Z)$, which is always of type $(1,1)$ as $h^{2,0}(Z)=0$,
a point on the critical locus thus determines the spectral data of
a holomorphic $E_8$ bundle on $Z$. Also, by changing the bounding
chains $\Gamma_j$, we see that $\psi_j$ is defined only modulo the
periods of $\Omega^{3,0}$ over the lattice $H_3(Z,{\bf Z})$, as
expected for a Chern-Simons-like functional. In the present
context, this is a reflection of the Hodge structure being of
mixed type.

In general we should probably not expect such nice divisors as
$R_1$ and $R_2$ above. We will use a different description. Recall
that despite the suggestive notation, $\Omega_0$ is not quite the
holomorphic four-form on $Y_t$, rather it is the $F^4$ part with
respect to the limiting mixed Hodge structure on $H^4(Y_t)$.
However, there is a sense in which it corresponds to a $(4,0)$
form. For $t\not = 0$ the holomorphic $(4,0)$ form is a section of
a line bundle ${\cal F}^4$ over the disk $\Delta_t$, which extends
over the origin. Geometrically such a section determines a
logarithmic $(4,0)$ form on the central fiber $Y_0$, up to
periodic identifications of the residues.

Now we can try to relate this to the discussion of section
\ref{CYGluing} by using Steenbrink's description of the limiting
mixed Hodge structure, which uses the logarithmic cohomology
$H^4_{\rm log}(Y_0)$. The limiting mixed Hodge structure
corresponds to the obvious Hodge filtration and a less obvious
weight filtration on $H^4_{\rm log}(Y_0)$, as discussed in
appendix \ref{MHS}. The $F^4$ part is generated by the logarithmic
$(4,0)$ form obtained above.

Let us look at the representing homomorphism for the extension
(\ref{W3W5Ext}), using this description of the limiting mixed
Hodge structure. We take an integral basis $\tilde f^j$ for
${\Gr}_4$, take its dual $\tilde f_j$, and lift to $f_j$ in the
dual of $\W_5/\W_3$. Geometrically, the $f_j$ correspond to cycles
in $H_4(W_1)$ or $H_4(W_2)$, up to a linear combination of
`vanishing cycles.' Then the relevant part of the representing
homomorphism is given by
\be \psi \ = \ \sum_j \tilde f^j \int_{f_j} \Omega_{Y_0}^{4,0}
\ee
The ambiguities by the vanishing cycles correspond to ambiguities
by the periods $\int_{e_i}\Omega_0$ discussed previously.

Thus we have
\be \psi_j \ =\ \int_{f_j} \Omega_0 \ =\ \int_{ f_j}
\Omega^{4,0}_{Y_0}\ee
But the expressions above are precisely the integrals we
previously encountered in section \ref{HolomCS}, up to the usual
ambiguity by $H_3(Z,{\bf Z})$. As we saw, we can rewrite them on
$Z$ as $\int_Z \Omega^{3,0} \wedge \omega_{CS}$, see eg. equation
(\ref{Transgression}). An alternative strategy is to note that
$\W_5/\W_3$ is dual to $\W_4$ with respect to the polarization, so
that we can equivalently study the extension class for $0 \to
\Gr_3 \to \W_4 \to \Gr_4 \to 0$.

In the final step, we want to lift $F^4 (\W_5/\W_3)$ to $F^4
\cap \W_5$. Equivalently since $F^4\cap \W_4=0$ and since we have
the exact sequence
\be 0 \ \to \ \W_4 \ \to \W_5 \ \to { \Gr}_5 \ \to \ 0 \ee
we can try to lift the extension class $\Ext^1_{\MHS}({\rm
Gr}_5,{\rm Gr}_4)$ to $\Ext^1_{\MHS}({\rm Gr}_5,W_4)$. From the
short exact sequence $0 \to {\rm Gr}_3 \to \W_4 \to {\rm Gr}_4 \to
0$, we get the long exact sequence
\be \ldots\to \Ext^0({\Gr}_5,{\Gr}_4) \to \Ext^1({\Gr}_5,{\Gr}_3)
\to \Ext^1({\Gr}_5,\W_4) \to \Ext^1({\Gr}_5,{\Gr}_4) \to 0 \ee
Here we use the fact (\cite{PetersSteenbrink}, proposition 3.35)
that $\Ext^p_{\MHS}$ vanishes when $p \geq 2$. So we see that
$\Ext^1_{\MHS}({\rm Gr}_5,{\rm Gr}_4)$ can always be lifted to
$\Ext^1_{\MHS}({\rm Gr}_5,\W_4)$, but the choice of lift is
parametrized by a class in $\Ext^1_{\MHS}({\rm Gr}_5,{\rm Gr}_3)$.
These extensions are indeed non-trivial. If $\Gr_5$ were
one-dimensional, this is essentially our earlier example, the
non-trivial extension of ${\bf Z}(0)$ by ${\bf Z}(1)$.

Let us try to write the representing homomorphism using the
logarithmic $(4,0)$ form. We take a three-cycle $d_k$ in
$H_3(Z,{\bf Z})$ and lift it to the dual of $\W_5$. We can
represent this by a pair of four-chains $g_k =(c_{1,k},c_{2,k})$
in $W_1$ and $W_2$, such that $\del c_{1,k} = - \del c_{2,k} =
d_k$. Then we have
\be \int_{g_k} \Omega_0 \ =\ \int_{g_k}\Omega^{4,0}_{Y_0} \ee
It is not completely clear how to write this as a simple
expression, but at least we can argue that the expression can be
localized on $Z$. To see this, suppose that we pick another lift
$\tilde c_{1,k}$ with $\del \tilde c_{1,k} = d_k$. Then $c_{1,k} -
\tilde c_{1,k}$ is a closed four-cycle on $W_1$. Therefore modulo
certain discrete ambiguities given by the periods
$\int_{f_j}\Omega_0$ (and actually also $\int_{e_i} \Omega_0)$,
the expression is independent of the extension of the cycle away
from $Z$. Furthermore we have already seen previously that these
ambiguities can be rewritten as expressions on $Z$, so the result
follows. We will write it informally as $\int_{g_k}
\Omega_0=\int_{d_k}\Phi$.

To summarize, let us collect all the pieces. The period map for
the nilpotent orbit is of the form
\be \Omega_0(t) \ = \ e^i \int_{d_i}\Omega^{3,0} + f^j\, \psi_j +
g^k\,\left( {1\over 2\pi i}\log(t) \int_{d_k} \Omega^{3,0} +
\int_{d_k} \Phi \right) \ee
Geometrically, we think of $\int_{e_i}\Omega(t)$ as periods over
the vanishing cycles, $\int_{f_j}\Omega(t)$ as periods over cycles
in $H_4(W_1) \oplus H_4(W_2)$, and $\int_{g_k}\Omega(t)$ as
periods over four-cycles which have to `cross the neck.'

For comparison with expressions predicted by dualities, we define
the variable
\be T = {1\over 2\pi i}\log(t) \ee
Then we have found that for any semi-stable degeneration $Y_t \to
Y_0 = W_1 \cup_Z W_2$ with $W_{1,2}$ log Calabi-Yau fourfolds,
the flux superpotential $W = {1\over 2\pi i}
\int_{Y_t}\Omega^{4,0}\wedge {\sf G}$ is of the asymptotic form
\be\label{WNilp} W\ = \ \int_Z \Omega^{3,0} \wedge H + \int_Z
\Omega^{3,0}\wedge \omega_{CS} +  T  \int_Z \Omega^{3,0}\wedge
\widetilde H + \int_Z \Phi \wedge \widetilde H + \cO(e^{2\pi i T})
\ee
Here the first term comes from $F^4 {\rm Gr}_5$; the second
part comes from the $F^4$ part of the extension of ${\rm Gr}_5$ by
${\rm Gr}_4$, and the last part comes from $\log(t) N $ acting on
$F^4 \cap W_5$ and from the $F^4$ part of the extension of
$\W_5/\W_3$ by $\W_3$. The exponential terms are corrections to
the nilpotent orbit, which should be interpreted as instanton
corrections. Note that the fluxes (or rather the combination $H +
\omega_{CS}$ as discussed in section \ref{HolomCS}) in general do
not take values in the full $H^3(Z,{\bf Z})$ lattice, but only in
${\rm coker}(d^3)^\perp$, the sublattice which is orthogonal to
the cokernel of $H^3(W_1) \oplus H^3(W_2) \to H^3(Z)$ with respect
to the intersection pairing.

Up to one issue which we discuss below, we see that this
reproduces known examples. In the heterotic $E_8\times E_8$ or
$SO(32)$ limits of $K3$ fibrations, the parameter $T$ is
identified as a K\"ahler modulus associated to the elliptic fiber
of the heterotic compactification, and the corrections are
interpreted as coming from worldsheet instantons whose
contribution is proportional to $\exp(2\pi i T)$. Similarly in the
Sen limit, $T$ is identified with axio-dilaton $\tau$ of type IIB
($\tau = i/g_s + a$), and the corrections are interpreted as
$D(-1)$-instantons, which do indeed give a contribution of the
form $\exp(2\pi i T)$. For previous work on periods in
heterotic/$F$-theory duality, see eg.
\cite{Clingher:2003ui,Grimm:2009ef,Jockers:2009ti,Aganagic:2009jq}.

Our analysis also gives further support for holography, i.e. the
claim that the theory can be thought of as supported on the
boundary $Z$. Moreover we see that the form of the Lagrangian is
almost completely fixed from very little input. The only thing we
don't know is exactly what CS theory we are dealing with. This
clearly depends on the concrete model under consideration. From
the example given earlier in the paper, it is clear that $SO(32)$
and $E_8\times E_8$ are not the only possible answers.

We have to make some remarks on the terms involving $\widetilde
H$. The $E_8\times E_8$ heterotic string has only a single
two-from field $B_{\mu\nu}$, so if we take the $E_8\times E_8$
limit then the superpotential is not expected to depend on a
second tensor field. This is not a contradiction. The expression
(\ref{WNilp}) really corresponds to all possible linear
combinations of the periods. $F$-theory places restrictions on the
allowed fluxes, namely they should be orthogonal to four-cycles
contained in in the base $B_3$ or of the form $\pi^{-1}(C)$, where
$C$ is a two-cycle in the base. This tends to place restrictions
on ${\sf G}$-fluxes coming from $H^4(W_1)$ and $H^4(W_2)$. If we
had allowed such fluxes, then for example the Chern-Simons theory
coming from the local $dP_9$-model would have a gauge group larger
than $E_8$, as was checked carefully in \cite{Donagi:2008ca}.
However the ${\sf G}$-fluxes which descend to $\widetilde H$ are
of a slightly different form. They are Poincar\'e dual to
four-cycles on $Y_t$ which asymptote to four-cycles on $Y_0$ of
the form $c_1+c_2$, where $c_{1,2}$ are four-chains in $W_{1,2}$
such that $\del c_1 = - \del c_2 = b \in H_3(Z,{\bf Z})$. It is
not clear to us why such fluxes would be disallowed, but we have
not checked it precisely.

\newsubsection{The K\"ahler potential}

\subseclabel{KPotential}

One may similarly compute the K\"ahler potential on the nilpotent
orbit. In the following, we will compare this with the $E_8\times
E_8$ limit. The K\"ahler potential is not protected, so it would
be remarkable if we would get an exact match. This is not the
case, but nevertheless we can match the main terms.

The K\"ahler potential in $F$-theory is of the following form:
\be {\cal K}_F \ = \ -\log\left[ \int_Y \Omega^{4,0} \wedge
\overline{ \Omega^{0,4}}\right]  -2 \log {\cal V}_{B_3} + {\cal
K}_{{\sf C}_3} \ee
We set the four-dimensional Planck scale $M_{Pl}$ equal to one.
Note that up to a constant, $\int_{Y_t}\Omega(t)\wedge
\overline{\Omega(t)}$ is simply $Q(\Omega,\overline{\Omega})$,
where $Q$ is the polarization on $H^4(Y_t)$. By general properties
of the monodromy, we have $Q(Na,Nb)=-Q(a,N^2b)=0$ for any forms
$a,b$. For simplicity, we will assume that the polarization is
simply the sum of the polarizations on the graded pieces. In
particular, $Q(\cdot, N \, \cdot)$ corresponds to the polarization
of the weight three Hodge structure on ${\rm Gr}_3 \simeq
H^3(Z,{\bf Z})$. Then we have
\be Q(\Omega(t),\overline{\Omega}(t)) \ \sim \ {\sf T}\, Q_3 + Q_4
\ee
where ${\sf T} = {\rm Im}({1\over 2\pi i}\log(t))$, $Q_3 = i\int_Z
\Omega^{3,0} \wedge \overline{ \Omega^{3,0}}$ and $Q_4$ is the
polarization on $\Gr_4$, evaluated on the image of $\Omega_0$
under the projection $H^4(Y_t) \to \Gr_4$. The K\"ahler potential
for the nilpotent orbit is then given by
\be\label{TKahlerPotential} {\cal K}_{\Omega(t)} \ \sim \
-\log\,\left[
 {\sf T}\, Q_3 + Q_4\right] \ \sim \ - \log ({\sf
T})-\log [Q_3] + {Q_4\over {\sf T}\, Q_3} + \cO( {\sf T}^{-2}) \ee
Furthermore using the isomorphism $\Gr_4 \to H^2_{\rm
v}(C_1)_\Lambda \oplus H^2_{\rm v}(C_2)_\Lambda$, we see that up
to a constant, $Q_4$ is mapped to the natural polarization on the
second cohomology of the spectral covers $C_1$ and $C_2$.

Now let us take a closer look at the K\"ahler moduli. In
heterotic/$F$-theory duality, the $F$-theory base $B_3$ is a ${\bf
P}^1$-fibration over a surface $B_2$. The volume is then given by
\be\label{B3Volume} {\cal V}_{B_3} \ = \ \half {\sf F}\, {\sf
B}^i\, {\sf B}^j d_{ij} + \half {\sf F}^2 {\sf B}^i d_{ij} n^j +
{1\over 6} {\sf F}^3 d_{ij} n^i n^j \ee
where ${\sf F}$ is the volume of the ${\bf P}^1$-fiber, ${\sf B}^i
d_{ij}$ are volumes of two-cycles in $B_2$, $n^i$ is the class of
the normal bundle of $B_2$ in $B_3$, and $d_{ij}$ is the
intersection form on $B_2$. We first need to rewrite this in terms
of the K\"ahler moduli, which are given by ${\sf S} = \del{\cal
V}_{B_3}/\del {\sf F}$ and ${\sf T}^i =d^{ij}\del{\cal
V}_{B_3}/\del {\sf B}^j$. If we further assume that $B_3 = B_2
\times {\bf P}^1$ is a direct product, then the terms involving
$n^i$ in (\ref{B3Volume}) are absent and we can easily rewrite
this in terms of the K\"ahler moduli as $ {\cal V}_{B_3} = {\sf
S}^{1/2} ({\sf T}^i {\sf T}^j d_{ij})^{1/2}$. Then we have
\be K_F \ \sim -\log({\sf S}) -\log({\sf T})-\log({\sf T}^i{\sf
T}^j d_{ij}) -\log [Q_3] + {Q_4\over {\sf T}\, Q_3} + {\cal
K}_{{\sf C}_3}  +\cO( {\sf T}^{-2}) \ee
where we omitted the additional terms in (\ref{TKahlerPotential})
and (\ref{B3Volume}). We also have $\del\delb {\cal K}_{{\sf C}_3}
\sim \int_Y \delta {\sf C}_3 \wedge * \delta {\sf C}_3$, but for
Calabi-Yau four-folds there are usually no moduli of the form
$\delta {\sf C}_3$ as generically $H^3(Y) = 0$.

On the $E_8\times E_8$ heterotic side, the K\"ahler potential is
of the form
\be {\cal K}_{\rm het} \ \sim \ -\log({\sf S}) -\log({\cal V}_Z)
-\log \left[i\int_Z \Omega^{3,0} \wedge
\overline{\Omega^{3,0}}\right] + {\cal K}_A \ee
where ${\cal K}_A$ denote the K\"ahler potential for the bundle
moduli. The volume ${\cal V}_Z$ can be written as
\be {\cal V}_Z\ =\ \half {\sf T}\,{\sf T}^i\,{\sf T}^j d_{ij} +
\half{\sf T}^2 {\sf T}^i d_{ij} k^j + {1\over 6} {\sf T}^3 d_{ij}
k^i k^j\ee
The potential for bundle moduli is harder to write, but we can
easily write the K\"ahler metric:
\be \del \delb {\cal K}_A\ \sim\ {1\over {\cal V}_Z}\int_Z {\rm
Tr}(\delta A \wedge \overline{\delta A})\wedge J \wedge J\ \sim\
{1\over {\sf T}}\int_{B_2} {\rm Tr}_E(\delta A_\nu \wedge
\overline{\delta A}_\nu)  + \cO({\sf T}^{-2}) \ee
Here $\delta A_\nu$ denotes the component of $\delta A$ along the
elliptic curve $E$, which correspond to normal bundle valued
moduli of the spectral cover. We can map them to $(2,0)$ forms on
the spectral cover by contracting with $\Omega^{3,0}_Z$. Thus we
can write
\be {1\over {\sf T}}\int_{B_2} {\rm Tr}_E(\delta A_\nu \wedge
\overline{\delta A}_\nu)\ \sim\ {1\over {\sf T}\,Q_3}\int_{B_2}
{\rm Tr}(\delta\Phi^{2,0} \wedge \overline{\delta \Phi^{2,0}}) \ee
Abelianizing $\delta \Phi^{2,0}$ we get $(2,0)$ forms on the
spectral cover, which get mapped to variations $\delta \Omega_0$
sitting in $\Gr_4$ under the cylinder map. So we see that this
describes the metric for complex structure moduli of the spectral
cover, and should be compared with $\del \delb Q_4/ {\sf T} Q_3$
on the $F$-theory side.

Comparing ${\cal K}_F$ with ${\cal K}_{\rm het}$, we see that the
main terms match qualitatively, without doing any K\"ahler
transformations. It might be interesting to do the comparison more
carefully, as this gives some insight into the subleading terms
which might be useful for the purpose of moduli stabilization.

In the IIb limit one can do a similar comparison. It is in fact
more straightforward as the potentials for the K\"ahler moduli can
be compared more easily.

\newsubsection{Euler character}

A number of additional properties of $Y_t$ can be deduced from the
degeneration limit. For example, the Euler character or more
generally Chern classes of $Y_t$ can be computed using
Chern-Schwartz-Macpherson classes in the degeneration limit. Here
we briefly consider the Euler character, which plays a role in
tadpole cancellation.  Recall that the cohomology groups of the
smooth fibers agree with the logarithmic de Rham cohomology groups
of the central fiber. Therefore if $Y_t \to Y_0=W_1 \cup_Z W_2$,
we have
\be\label{VerdierChi} \chi(Y_t) \ = \ \sum_k (-1)^k\, H^k_{\rm
log}(Y_0) \ = \ \chi(W_1) + \chi(W_2) - 2\chi(Z) \ee
Note that this differs from the topological Euler character of
$Y_0$. The extra factor of $-\chi(Z)$ comes from the logarithmic
forms that have a pole along $Z$, see equation (\ref{LogResLES}).

We consider the following application. Suppose that we use the
Tate form of the Weierstrass fibration to engineer an elliptic
fibration with interesting singularities along a divisor $S
\subseteq B_3$. The Euler character of the resolved $Y$ will
depend on the singularities, and one may ask how it changes when
we put different singularities along $S$. When $Y$ has a heterotic
dual, we can deduce the answer from heterotic/$F$-theory duality.
It was observed in section 2.3 of \cite{Blumenhagen:2009yv} that
the same formula appears to hold more generally, when there is no
heterotic dual (i.e. $Y$ does not admit a K3 fibration).

From the perspective of the present paper, this observation is a
corollary of the result (\ref{VerdierChi}) together with the
degeneration limit of section \ref{localglobal}. When the
singularities along $S$ fit in $E_8$, we can use our degeneration
limit to move the singularities to $W_1$. But we already know
$\chi(W_1)-\chi(Z)$ from heterotic/$F$-theory duality. Namely, it
should agree with $\int_S {\rm ch}_2(V) - \half {\rm ch_2}(TZ)$,
which was computed in \cite{Friedman:1997yq}, upon setting $\gamma
= 0$ in their formulae (as this part gets mapped to the ${\sf G}$-flux).

\newsubsection{Bulk versus boundary}

We would like to end with a few more comments.

Perhaps one of the most intriguing aspects of this paper is the
interplay between the theory in the bulk and the theory on the
boundary. For the $E_8\times E_8$ and $SO(32)$ degeneration limits
of the $K3$-surface, it has long been known that the heterotic
theory may be thought of as living on the boundary (the normal
crossing divisor), but given the lack of examples it was less than
clear how general this is. In sections \ref{CYGluing} and
\ref{VHS} we have given arguments that this should be true in more
general degeneration limits, and in \cite{CDW} we give another
concrete example of this phenomenon. It would be interesting to
understand the general boundary theories in more detail.

Kawamata/Namikawa also allow the normal crossing divisor to have
multiple components $Z_i$ which intersect. This leads us to
consider defects in the boundary theory. The defect theories
should be very interesting. As indicated in section
\ref{CYGluing}, we expect to find a kind of holomorphic
generalization of WZW models on such defects. One should be able
to deduce some of its properties from an analysis of the limiting
mixed Hodge structure, similar to sections \ref{LimMHS} and
\ref{WAsymptotics}.

Log geometry plays a central role in the mirror symmetry program
of Gross and Siebert \cite{GSMirror1}, and their philosophy of
gluing Calabi-Yau manifolds from more elementary pieces is similar
to the one in this paper, although their singular Calabi-Yau
spaces are not necessarily of normal crossing type. It would be very
interesting to understand the holographic aspects of their work.

Finally, the superpotential still receives corrections that are
non-perturbative in the K\"ahler moduli, due to Euclidean brane
configurations. In type II these come from $D$-instantons, in
$F$-theory they come from $M5$-instantons and in the heterotic
string they come from worldsheet instantons and $NS5$-instantons.
They are all related through heterotic/$F$-theory/type II duality.
It is hard to escape the impression that there must be some
underlying variation of Hodge structure problem that encodes all
instanton corrections to the superpotential in $4d$
compactifications with $N=1$ supersymmetry. It would be extremely
interesting to elucidate this.

\medskip

\noindent {\it Acknowledgements}: We are grateful to P.~Aluffi,
A.~Clingher, J.~Koll\'ar, Y.~Kawamata, T.~Pantev, R.~Thomas,
B.~Siebert, and E.~Witten for discussions and correspondence
related to this project. RD acknowledges partial support by NSF
grants 0908487 and 0636606.  SK acknowledges partial support by
NSF grants DMS-05-55678 and DMS-12-01089 in addition to the hospitality
of the University of Pennsylvania.
The research of MW is supported by a
Heisenberg grant from the DFG.

\newpage

\appendix

\renewcommand{\newsection}[1]{
\addtocounter{section}{1} \setcounter{equation}{0}
\setcounter{subsection}{0} \addcontentsline{toc}{section}{\protect
\numberline{\Alph{section}}{{\rm #1}}} \vglue .6cm \pagebreak[3]
\noindent{\bf Appendix {\Alph{section}}:
#1}\nopagebreak[4]\par\vskip .3cm}

\newsection{Mixed Hodge structures and monodromy}

\seclabel{MHS}

In this appendix we briefly review some aspects of Hodge theory
that will help us understand a semi-stable degeneration limit.
There are many books and reviews available on this material, see
for example \cite{GriffTranscendental,VoisinHodge} and
\cite{PetersSteenbrink}.

Recall that on a compact K\"ahler manifold $X$ of complex
dimension $n$, the cohomology groups admit a Hodge decomposition
\be H^k(X,{\bf C})\ =\ \sum_{p+q=k} H^{p,q}(X). \ee
Let us denote the K\"ahler form of $X$ by $J$, which we may take
to be a class in $H^2(X,{\bf Z})$. The primitive part of the
cohomology, denoted by $PH^k(X,{\bf Z})$, is defined to be the
kernel of $\wedge J^{n-k+1}:H^k(X,{\bf Z}) \to H^{2n-k+2}(X,{\bf
Z})$. Then we have a further decomposition of the cohomology, the
Lefschetz decomposition:
\be H^k(X,{\bf Q}) \ = \ \bigoplus_m J^m \cdot PH^{k-2m}(X,{\bf
Q}) \ee
Since $J$ is of type $(1,1)$, the interesting part of the
information is contained in the primitive part of the cohomology,
and we will often implicitly assume that we are restricting to the
primitive part.

We further have a polarization, i.e. a non-degenerate bilinear
form
\be Q(a,b) \ = \ (-1)^{k(k-1)/2}\int_X a\wedge b \wedge J^{n-k}\
\ee
which is symmetric for $k$ even and anti-symmetric for $k$ odd. It
satisfies the following two relations, which are called the
Hodge-Riemann bilinear relations: we have $Q(H^{p,q},H^{p',q'}) =
0$ unless $p+p'=q+q'=k$, and $i^{p-q} Q(a,\bar a) \geq 0$ if $a\in
PH^{p,q}$.

We will be interested in period integrals on $X$. The periods
themselves depend on some redundant information, as we have to
make a non-canonical choice of basis. A more invariant way to
state it is that we are interested in how the Hodge decomposition
$\bigoplus H^{p,q}(X)$ of $H^k(X,{\bf C})$ varies in families.
However the $H^{p,q}(X)$ individually do not vary holomorphically
even if $X$ lives in a holomorphic family.

Let $\pi_{{\cal X}}: {\cal X} \to B$ denote a smooth holomorphic family,
with fibers $X_t$ for $t \in B$. On each fiber $X_t$ we have the
cohomology lattice $H^k(X_t,{\bf Z})$. As we vary $t \in B$, these
lattices fit together in a local system ${\cal L}^k$ over $B$.
Mathematically, we have
${\cal L}^k =R^k\pi_{{\cal X}*}{\bf Z}$. Similarly, for each fiber
$X_t$ we have a vector space $H^k(X_t,{\bf C})$. As we vary $t \in
B$, these vector spaces fit together in a vector bundle ${\cal
H}^k\to B$, which is just the complexification ${\cal H}^k = {\cal
L}^k\otimes \cO_B$. The vector bundle ${\cal H}^k\to B$ is flat
and holomorphic. It has a natural flat Gauss-Manin connection
$\nabla$, obtained by requiring that the local sections induced from
local sections of ${\cal L}^k$ are flat.

Now we may put a decreasing filtration ${\cal F}^\bullet$ on the
bundle ${\cal H}^k$, by putting the following decreasing
filtration on each of the fibers:
\be 0 \subseteq F^k \subseteq F^{k-1} \subseteq \ldots \subseteq
F^0 =H^k(X_t,{\bf C}), \qquad F^p \ \equiv \  \bigoplus_{q\geq p}
H^{q,k-q}(X_t) \ee
Note that we may recover $H^{p,q}$ as $H^{p,q}(X_t) =F^p \cap
{\overline{ F}}^q$, so we did not lose any information by
focusing on the filtration. Although the $H^{p,q}$ generally do
not vary holomorphically, the subbundles ${\cal F}^p \to B$ with
fibers $F^p \subset H^k(X_t,{\bf C})$ do vary holomorphically.

The filtration satisfies the following properties. Let us denote
$H_{\bf Z} = H^k(X_t,{\bf Z})$. Then $F^\bullet$ is a $k$-step
decreasing filtration of $H_{\bf C} = H_{\bf Z} \otimes {\bf C}$,
such that $H_{\bf C} \cong F^p \oplus {\overline{F}}^{k-p+1}$ for
all $p$. We also have a bilinear form $ Q(\cdot,\cdot)$ with the
properies listed previously. As we vary $t \in B$, this yields a
local system ${\cal L}^k$ over $B$ and a decreasing filtration ${\cal
F}^\bullet$ of ${\cal H}^k={\cal L}^k\otimes{\cal O}_B$,
such that each ${\cal F}^p$
is a holomorphic vector bundle. Furthermore, the ${\cal F}^p$
behave nicely with respect to the Gauss-Manin connection, namely
they satisfy Griffiths transversality
\be \nabla: {\cal F}^p \to {\cal F}^{p+1} \otimes \Omega^1_B \ee
The data $(B,{\cal L}^k, {\cal F}^\bullet, Q(\cdot,\cdot),\nabla)$
satisfying the properties discussed above defines a polarized
variation of Hodge structure of weight $k$.

A filtration in a fixed vector space
is classified by a point in the flag manifold ${\bf
F}(n_k,..,n_0)$, where $n_p = \dim F^p$.  After pullback to the universal
cover $\rho:\tilde{B}\to B$, the bundle $\rho^*{\cal H}^k$ becomes trivial.
Fixing a reference point in $\tilde{B}$ over a point $t\in B$, the
Gauss-Manin connection gives an explicit isomorphism
$\sigma:\rho^*{\cal H}^k\simeq H^k(X_t,{\bf Z})\otimes{\cal O}_{\tilde{B}}$.
Using this isomorphism we get a holomorphic period mapping
\be \Pi:\,\tilde{B}\ \to\ {\bf F}(n_k,\ldots,n_0), \ee
by associating to each $s\in \tilde{B}$ the flag ${\cal F}^{\bullet}_{\rho(s)}$,
identified with a flag in $H^k(X_t,{\bf Z})$ via $\sigma$.
If the Hodge structure is polarized, its image is
contained in the {\em period domain\/}, which consists of the flags
which satisfy the
Hodge-Riemann bilinear relations, i.e.\ $Q(F^p,F^{k-p+1})=0$, and
$i^{p-q} Q(a,\bar a) \geq 0$ if $a\in
PH^{p,q}$.

By introducing a basis of integral cycles of $X$, we can write the
mapping $\Pi$ in coordinate form in terms of period integrals, but
this explicit form obviously depends on a non-canonical choices.
Continuing to fix our reference point $t\in B$,
we pick a basis $e_i \in H_k(X_{t},{\bf Z})$ and its dual basis $e^i \in
H^k(X_{t},{\bf Z})$. We use the Gauss-Manin connection to
extend this basis away from $t$ by parallel
transport.  Globally, these bases becomes multi-valued over $B$ but
single-valued over $\tilde{B}$.
We also need to introduce a basis for the ${\cal F}^p$.
Since the ${\cal F}^p$ are holomorphic vector bundles, locally on
an open subset $U \subset B$ we can trivialize these bundles by
choosing a holomorphic frame. It is convenient to adapt this basis
to the filtration, i.e. we require $\{\omega_1,.., \omega_{n_k}\}$
to project to a holomorphic frame for ${\cal F}^k$, $\{
\omega_{n_k + 1} ,..,\omega_{n_{k-1}}\}$ to project to a frame for
${\cal F}^{k-1}/{\cal F}^k$, etc. Then the period mapping
associates to each $t \in U$ the explicit flag
\be 0\ \subseteq\ \vev{ x_{i,1} e^i ,\ldots,x_{i,n_k} e^i }\
\subseteq\ \vev{x_{i,1} e^i,\ldots,x_{i,n_{k-1}} e^i }\ \subseteq\
\ldots\ \subseteq\
 H^k(X_t,{\bf C}) \ee
Here
\be x_{i,r} \ = \ \int_{e_i} \omega_r \ee
and $\vev{\ldots}$ denotes the linear subspace of $H^k(X_t,{\bf
C})$ generated by the vectors between the brackets.

We are interested in degenerating a Hodge structure with some
weight $k$. It turns out that in the limit the Hodge structure is
not necessarily of pure type, but may contain pieces with higher
or lower weight than $k$. This leads to some extra structure in
the form of an increasing weight filtration $W_{\bullet}$ of
$H^k(X,{\bf Q})$:
\be 0\ \subseteq \ W_0\ \subseteq\ \ldots\ \subseteq\ W_{i}\
\subseteq\ \ldots\ \subseteq\ W_{2k}=H^k(X,{\bf Q})
 \ee
The filtrations $W_\bullet$ and $F^\bullet$ define
a {\em mixed Hodge structure\/} when the weight filtration is
compatible with the Hodge filtration ${\cal F}^\bullet$, in the
sense that the induced filtrations ${\cal F}_i^\bullet$ on the graded pieces
\be {\rm Gr}_i^W \ = \ W_i/W_{i-1} \ee
satisfy the axioms of
a Hodge structure of weight $i$. By duality between homology
and cohomology, we get an induced filtration on the homology, whose
weights are the negatives of the weights on cohomology. Here we must use the
conventional duality $H_* = \Hom(H^*,{\bf Z})$, not Poincar\'e
duality, which tends to fail on singular spaces.

It has been shown by Deligne that the cohomology of any variety
$X$ canonically carries such a mixed Hodge structure. There are
two basic cases: if $X$ is complete but possibly singular, then
$W_i =H^k(X,{\bf Q})$ for $i \geq k$. If on the other hand $X$ is
smooth but not necessarily complete, then $W_i = 0$ for $i< k$.

Let us now suppose that we have a one-parameter semi-stable
degeneration, i.e. we have a one-parameter family $\pi_{{\cal X}}:
{\cal X} \to \Delta$ with fibers $X_t$, such that the total space
is smooth and the central fiber is reduced and has at most normal
crossing singularities. There are two mixed Hodge structures that
we can associate to this set-up. The first is the mixed Hodge
structure of the central fiber $X_0$, and the second comes from
the monodromy weight filtration. Let us discuss each of these in
turn.

First we consider the mixed Hodge structure of the central fiber.
Recall that $X_0$ is a complete and reduced variety with only
normal crossing singularities. The weight filtration in this case
can be read of from the Mayer-Vietoris spectral sequence. Let us
suppose that $X_0$ decomposes into irreducible pieces as $X_0 =
D_1 \cup \ldots \cup D_N$. We set $D_{i_1\ldots i_q} = D_{i_1}
\cap \ldots \cap D_{i_q}$ and define the codimension $q$ stratum
as
\be X_0^{[p]} \ = \ \coprod_{1 \leq i_1 <\ldots <i_p\leq N}
D_{\,i_1 \ldots i_p} \ee
where $\amalg$ denotes disjoint union. Now let us define
\be E_0^{p,q}\ \equiv\ {\cal A}^q(X_0^{[p+1]})\ee
where ${\cal A}^\bullet$ denotes De Rham cohomology. As usual, we
have the exterior derivative $d : E_0^{p,q} \to E_0^{p,q+1}$. In
addition, we have a second differential $\delta:E_0^{p,q} \to
E_0^{p+1,q}$, which is the restriction map given by
\be \delta \phi(D_{i_1 \ldots i_{p+1}})\ =\ \sum_a (-1)^a\,
\phi(D_{i_0\ldots\hat i_a \ldots i_{p+1}}) \ee
Then we have $d^2=\delta^2=d\delta + \delta d = 0$, so the data
$(E_0^{\bullet,\bullet},d,\delta)$ yields a double complex. It can
be shown that this spectral sequence converges to $H^k(X_0,{\bf
R})$.

Now we can use this to define a filtration:
\be\label{NCrossWFiltration} W_i \ =\ \bigoplus_{r\leq i}
E_0^{\bullet,r} \ee
This descends to a weight filtration on $H^k(X_0,{\bf R})$.
Similarly we define a Hodge filtration as $F^p = \bigoplus_{r,s}
F^pE_0^{r,s}$, which descends to $H^k(X_0,{\bf C})$.

Let us discuss this in more detail for the main case of interest
in this paper, where we have $X_0 = W_1 \cup_Z W_2$. Then we have
\be E_1^{0,k}=H^k(W_1)\oplus H^k(W_2),\qquad E_1^{1,k}=H^k(Z),\ee
with all other terms vanishing. The non-zero differentials are the
restriction maps
\be d_1^{0,k}:H^k(W_1)\oplus H^k(W_2)\ \to\ H^k(Z) \ee
and the spectral sequence degenerates at $E_2$.  This gives a
short exact sequence
\be 0\ \to\ {\rm coker}(d_1^{0,k-1})\ \to\  H^k(X_0)\  \to\  {\rm
ker}(d_1^{0,k})\ \to\ 0 \ee
which corresponds to the weight filtration
\be 0 \ \subseteq\ W_{k-1} \ \subseteq  \ W_{k} = H^k(X_0,{\bf R})
\ee
with $\Gr_{k}H^k(X_0)\cong\ker(d_1^{0,k})$ and
$\Gr_{k-1}H^k(X_0)\cong{\rm coker}(d_1^{0,k-1})$. In fact, in this
example it is simpler to read off the weight filtration directly
from the long exact sequence version of Mayer-Vietoris:
\be\label{MVLES} \ldots\ \to \ H^{k-1}(Z) \ \to \ H^k(W_1 \cup_Z
W_2) \ \to \ H^k(W_1) \oplus H^k(W_2) \ \to \ H^k(Z) \ \to \
\ldots \ee
We see that the graded pieces ${\rm Gr}_{k-1}$ and ${\rm Gr}_k$
should be thought of as a quotient or a subspace of the cohomology
groups of $Z$ and $W_1 \amalg W_2$ respectively. These are smooth
and complete spaces, which is why they each carry a pure Hodge
structure. This is in some sense the general picture: we find a
suitable stratification to construct a weight filtration, so that
the graded pieces can be though of as the Hodge structure of a
complete non-singular variety, and hence carry a pure Hodge
structure.

The second mixed Hodge structure associated to our semi-stable
degeneration comes from the monodromy. As we circle around the
origin $t=0$, the cycles in $H_k(X_t,{\bf Z})$ get rearranged.
This is expressed as an automorphism $M$ of $H^k(X_t,{\bf Z})$
called the monodromy transformation. It is known for our semi-stable
degeneration that the monodromy is unipotent. We also
introduce the logarithm $N = \log(M)$, which is nilpotent since $M$
is unipotent.

The operator $N$ should be thought of as an analogue of the
Lefschetz operator (cup product with the K\"ahler class). As such
it gives rise to a kind of Lefschetz decomposition on
$H^k(X_t,{\bf Q})$, which in turn gives the weight filtration. The
monodromy weight filtration is completely determined by the
following two properties. The log of the monodromy matrix has
degree $-2$ (i.e. $N\cdot W_i \subset W_{i-2}$), and $N^j$ gives
an isomorphism $N^j: Gr_{k+j} \to Gr_{k-j}$. For example, if
$N^2=0$ but $N\not = 0$, then the monodromy weight filtration
simply takes the form
\be 0\ \subseteq\ {\rm im}(N)\ \subseteq\ {\rm ker}(N)\ \subseteq\
H^k(X_t,{\bf Q})\quad \ee
i.e. we have
\be W_{k-1} = {\rm im}(N), \qquad W_k = {\rm ker}(N), \qquad
W_{k+1} = H^k(X_t,{\bf Q}) \ee
In the polarized case, we further have $Q(Na,b) + Q(a,Nb) = 0$.
The polarization $Q$ of $H^k(X_t,{\bf Z})$ yields a natural
polarization on the graded pieces, i.e.\
$Q(\cdot, N^j\, \cdot)$ on $\Gr_{k+j}$.
For example in the case $N^2 =
0$, $N\not = 0$ above, we have $(x,y) \to Q(x,Ny)$ for $x,y \in
H^k(X_t,{\bf Q})/{\rm ker}(N)$, and we have $(x,y) \to Q(\tilde
x,y)$ for  $x,y \in {\rm im}(N)$, where $\tilde x$ is a lift from
${\rm im}(N)$ to $H^k(X_t,{\bf Q})$ with $N\tilde x = x$.

So we have now used the monodromy to put a weight filtration on
$H^k(X_t,{\bf Q})$. But this weight filtration is not compatible
with the Hodge filtration on $X_t$, because $X_t$ is smooth and so
the Hodge structure is of pure type. However it has been shown by
Schmid that the limiting Hodge filtration
\be {\cal F}_\infty^\bullet \ = \ \lim_{t\to 0}\ e^{-{1\over 2\pi
i}\log(t) N} {\cal F}^\bullet \ee
exists, and together with the monodromy weight filtration it does
define a mixed Hodge structure. It is called the limiting mixed
Hodge structure.

There is an alternative description of the limiting mixed Hodge
structure due to Steenbrink \cite{Steenbrink76}, which uses the
logarithmic cohomology groups of $X_0$. We discuss this a bit more
explicitly for $X_0 = W_1 \cup_Z W_2$. As in section
\ref{SemiStable} we use the short-hand notation $H^k_{\rm
log}(X_0) = {\mathbb H}^k(X_0, \Omega_{X_0}^\bullet(\log Z)) $.
The spectral sequence for ${\mathbb H}^k$ degenerates at the $E_1$
term, so the Hodge decomposition is the obvious one, $H^k_{\rm
log}(X_0) = \sum_{k = p+q}H^p(X_0, \Omega^q(\log Z))$. The weight
filtration can be obtained as follows. We have the short exact
sequence
\be 0 \ \to \ \Omega_{X_0}^p \ \to\ \Omega_{X_0}^p(\log Z) \
\mathop{\to}^{\rm res} \ \Omega_Z^{p-1} \ \to \ 0 \ee
where `res' denotes the Poincar\'e residue map. This gives the
long exact sequence
\be\label{LogResLES} \ldots\ \to \ H^{k-2}(Z)\ \to \ H^k(X_0) \
\to \ H^k_{\rm log}(X_0) \ \to \ H^{k-1}(Z) \ \to \ \ldots \ee
where all the maps are compatible with the Hodge structure, and
the coboundary map $H^{k-2}(Z) \to H^k(X_0)$ is a Gysin map. We
find that $\W_{k+1} = H^k_{\rm log}(X_0)$ and $\W_k =
H^k(X_0)/{\rm Im}(H^{k-2}(Z))$. From the Mayer-Vietoris sequence
for $X_0$ (\ref{MVLES}) we further get $\W_{k-1} = {\rm
coker}({\sf d}^{k-1})$ as before.

Given a limiting mixed Hodge structure, we can define the
nilpotent orbit as
\be \cO\ =\ \{ e^{{1\over 2\pi i}\log(t)N}{\cal F}_\infty^\bullet
\, | \, t \in \Delta \} \ee
The nilpotent orbit may be thought of as a kind of perturbative
approximation to the periods in the limit $t \to 0$. The precise
statement is the content of the nilpotent orbit theorem of Schmid
\cite{SchmidOrbits}. As before let
$\Pi$ denote the period mapping from the upper half plane
(the universal cover of the punctured disc) to
the period domain of the Hodge structure.  Define
$T = {1\over 2\pi i} \log(t)$, the usual coordinate on the upper half plane.
Then for any
reasonable metric $d$ on the period domain, there are
constants $A,B$ such that for ${\rm Im}(T) \geq A > 0$ we have
\be d(\Pi(\cO),\Pi({\cal F}))\ <\ {\rm Im}(T)^B e^{-2\pi\, {\rm
Im}(T)}. \ee
In other words, the corrections die out exponentially fast as
${\rm Im}(T) \to \infty$.

So in order to understand the behaviour of the periods as $t\to
0$, it is of interest to know the nilpotent orbit, or equivalently
the underlying limiting mixed Hodge structure. Note that the
limiting mixed Hodge structure does not sit on the usual
cohomology groups of $X_0$ (although we saw that it does sit on
the logarithmic version of the cohomology groups of $X_0$). The
usual cohomology groups of the central fiber do carry a canonical
mixed Hodge structure, but this differs from the limiting mixed
Hodge structure. Nevertheless there exists a close relation
between them. The precise relation is described by the
Clemens-Schmid long exact sequence, which we discuss next.

Let us denote by $H^k_{\rm lim}(X_t)$ the cohomology group of
$X_t$ equipped with the limiting weight and Hodge filtrations. The
Clemens-Schmid sequence is the following long exact sequence:
\be \ldots\ \mathop{\to}^\alpha \ H^k(X_0)\ \mathop{\to}^{i^*} \
H^k_{\rm lim}(X_t)\ \mathop{\to}^N\ H^k_{\rm lim}(X_t)\
\mathop{\to}^\beta \ H_{2n-k}(X_0)\ \mathop{\to}^\alpha\
H^{k+2}(X_0)\ \mathop{\to}^{i^*} \ \ldots \ee
The maps are as follows. The map $X_0 \hookrightarrow {\cal X}$ is
a homotopy equivalence, so we have $H^*(X_0)\cong H^*({\cal X})$.
The inclusion $i:X_t \hookrightarrow {\cal X}$ yields the
restriction map $i^*:H^k({\cal X}) \to H^k(X_t)$ on cohomology.
The map $\beta$ is simply the composition of Poincar\'e duality on
$X_t$, $H^k(X_t) \to H_{2n-k}(X_t)$, followed by the inclusion
$i_*:H_{2n-k}(X_t) \to H_{2n-k}({\cal X}) \cong H_{2n-k}(X_0)$. Of
course Poincar\'e duality on a smooth space is an isomorphism, so
the interesting part of this map is $i_*$. Finally, the map
$\alpha$ is the composition of Poincar\'e duality on the total
space ${\cal X}$, $H_{2n-k}({\cal X}) \to H^{k+2}({\cal
X},\del{\cal X})$, followed by the natural map $H^{k+2}({\cal
X},\del{\cal X}) \to H^{k+2}({\cal X})$. All the maps are
morphisms of mixed Hodge structures, i.e. morphisms of the
underlying lattices which preserves both the Hodge and the weight
filtration, with a degree shift on the Hodge type: $\alpha$ has degree
$(n+1,n+1)$, $i^*$ has degree $(0,0)$, $N$ has degree $(-1,-1)$,
and $\beta$ has degree $(-n,-n)$.

To summarize, given a semi-stable degeneration, we can figure out
the behaviour of the periods by taking the following steps: we
find the mixed Hodge structure on the central fiber, which can be
deduced from a stratification as in equation
(\ref{NCrossWFiltration}); then we use Clemens-Schmid or
alternatively the logarithmic cohomology groups to figure out the
limiting mixed Hodge structure and the nilpotent orbit; and
finally we use the nilpotent orbit theorem to conclude that the
latter is a good approximation to the periods.

\newsection{On the value of the holomorphic Chern-Simons action}

We discussed some aspects of holomorphic Chern-Simons theory in
section \ref{CYGluing}. Here we would like to would like to argue
that the value of the holomorphic Chern-Simons action evaluates to
zero for holomorphic bundles that admit a spectral construction,
modulo the usual periodic identifications.  We were motivated to
write this appendix after a question raised by E. Witten regarding
the Chern-Simons contribution to the gravitino mass in heterotic
string compactifications, $m_{3/2} = e^{{\cal K}/2}|W_0|$.

From the perspective of heterotic/$F$-theory duality, the reason
for the vanishing is very simple. The superpotential of the dual
$F$-theory compactification is given by
\be W_{F} \ = \ {1\over 2\pi}\int_Y \Omega^{4,0} \wedge {\sf G}
\ee
The part of the ${\sf G}$-flux that comes from the heterotic
bundle data is of Hodge type $(2,2)$. The superpotential
automatically evaluates to zero on such fluxes. The same argument
implies that the $D7$-brane superpotential does not contribute to
the gravitino mass in IIb vacua that can be lifted to $F$-theory.

This argument not very direct and perhaps somewhat unsatisfactory.
We will try to give a more conceptual argument below. It relies
however on a construction which is still partly conjectural.

Let us consider a holomorphic $G$-bundle on an elliptically
fibered Calabi-Yau three-fold $Z$ with section. We assume that the
bundle is semi-stable on generic fibers. Let us temporarily focus
our attention on a single elliptic fiber $E$. For every
semi-stable $G$-bundle on $E$, we can construct a rational surface
$R$ containing $E$ as the boundary, and a canonical $G$-bundle $V$
on $R$, such that the restriction $V|_E$ recovers our $G$-bundle
on $E$. This is relatively well-known when $G$ is an exceptional
group, in which case $R$ is a del Pezzo surface
\cite{Friedman:1997yq,FM}, but it can in fact be done for any gauge
group $G$ (see eg. \cite{LZ1}). For the case $G=SO(32)$, this
plays an important role in the $SO(32)$ limit of $F$-theory
\cite{CDW}.

We can now consider a relative version of this construction, by
fibering over a base $B$. We will want to consider the case $\pi:
Z \to B$ where $Z$ is our Calabi-Yau three-fold. Then we should
get a holomorphic bundle $V$ on a log four-fold $(Y,Z)$ which
restricts to our original bundle on $Z$. Furthermore, we expect
that the twisting data is given by
\be {{\sf G}\over 2\pi} \ = \ p_1(V) - \half p_1(TY(-{\log}Z)) \ee
These statements have not been fully shown. The current method is
to do a Fourier-Mukai transform along the elliptic fibers and then
use a cylinder mapping to construct $Y$ together with a Deligne
cohomology class, and this has not been carried out for all groups
$G$. When it has, it has not been shown mathematically that the
resulting $(2,2)$ flux can be interpreted as above. But we believe
that these constructions can be carried out.

Assuming this, we see that our holomorphic bundle on $Z$ can
actually be extended to a holomorphic bundle on $(Y,Z)$. By the
transgression argument discussed previously in section
\ref{CYGluing}, we can now directly show that the holomorphic
Chern-Simons action can be rewritten as
\be W_{CS} \ = \ {1\over 2\pi}\int_Y \Omega^{4,0} \wedge ({\sf
G}-{\sf G}_0) \ee
with the ${\sf G}$-flux above. Here we added a reference flux
${\sf G}_0$ such that ${\sf G} - {\sf G}_0$ restricts to zero in
$H^4(Z)$. Writing ${\sf G}_0^{0,4} = \delb \omega$, it actually
depends only on $\omega|_Z$. The dependence on a basepoint can 
be cancelled when we consider the full Chern-Simons action, by 
tadpole cancellation in the heterotic string, or because the 
relevant ${\sf G}$-flux lives in ${\rm ker}({\sf d}^4)$ from the 
point of view of section \ref{VHS}.

Now since our extended bundle is holomorphic, ${\sf G}$ is of type
$(2,2)$, and so the holomorphic Chern-Simons action evaluates to zero.
Conceptually this is the holomorphic analogue of a statement in
ordinary Chern-Simons theory, which says that the action
automatically vanishes if a flat connection on a real
three-manifold $M_3$ can be extended to a flat connection on $M_4$
with $M_3 = \del M_4$, as one would have ${\rm Tr}(F \wedge F) =
0$.

The argument may be applied to other examples. Consider for instance
a quintic three-fold. It is precisely the log boundary of ${\bf P}^4$.
So any bundle on the quintic that extends to a holomorphic bundle
on ${\bf P}^4$, or is constructed from such, should have zero Chern-Simons
invariant.

One may wonder at this point if one can produce examples with non-zero
Chern-Simons invariant, and hence non-zero gravitino mass. At least if
one ignores the issue of stability, then
it is not hard to do so. Let us simply take a configuration of curves $C_i$
on the quintic, and wrap heterotic five-branes on them with multiplicity
$n^i$, such that $\sum_i n^i [C_i]$ is a trivial class in homology.
Then by classic work of Griffiths and Clemens (see eg. \cite{VoisinHodge}),
there are configurations whose Abel-Jacobi map is non-zero,
for example the difference of two lines. By a theorem of Voisin,
such examples exist on all families of Calabi-Yau three-folds.
By varying moduli we can take
the projection on the $(0,3)$ component to be non-zero. However such a
configuration is clearly not stable as some of the $n^i$ have to be taken
negative. Producing stable examples appears to be harder.

\end{document}